\documentclass[12pt]{article}

\usepackage{graphics,graphicx,fullpage,natbib,multirow}
\usepackage{amsmath,amssymb,verbatim,epsfig}
\usepackage[dvipsnames,usenames]{color}

\newtheorem{defin}{\bf Definition}

\def\ga{\mbox{Ga}}

\def\gga{\mbox{GG}}
\def\sgg{\mbox{SGG}}

\def\be{\mbox{Be}}

\def\no{\mbox{N}}

\def\un{\mbox{Un}}

\def\gpd{\mbox{GPD}}
\def\E{\mbox{E}}
\def\V{\mbox{Var}}

\def\P{\mbox{P}}

\def\data{\mbox{data}}

\def\bx{{\bf x}}
\def\by{{\bf y}}

\def\bX{{\bf X}}
\def\bY{{\bf Y}}

\def\simind{\stackrel{\mbox{\scriptsize{ind}}}{\sim}}
\def\simiid{\stackrel{\mbox{\scriptsize{iid}}}{\sim}}

\newcommand{\btheta}{\boldsymbol{\theta}}

\newcommand{\PD}{\mathcal{PD}}

\newcommand{\RB}{\mathbb{R}}

\begin{document}

\baselineskip=24pt

\title{\bf Modelling heavy tail data with bayesian nonparametric mixtures}
\author{{\sc Luis E. Nieto-Barajas} \\[2mm]
{\sl Department of Statistics, ITAM, Mexico} \\[2mm]
{\small {\tt luis.nieto@itam.mx}} \\}
\date{}
\maketitle

\begin{abstract}
In the study of heavy tail data, several models have been introduced. If the interest is in the tail of the distribution, block maxima or excess over thresholds are the typical approaches, wasting relevant information in the bulk of the data. To avoid this, mixture models for the body (below the threshold) and the tail (above the threshold) are proposed. In this paper, we exploit the richness of nonparametric mixture models to model heavy tail data. We specifically consider mixtures of shifted gamma-gamma distributions with four parameters and a Poisson-Dirichlet process as a mixing distribution. One of these parameters is associated with the tail. By studying the posterior distribution of the tail parameter, we are able to assess the tail heaviness for each component. We develop an efficient MCMC method with adapting Metropolis-Hastings steps to obtain posterior inference and illustrate with simulated and real datasets. 
\end{abstract}

\vspace{0.2in} \noindent {\sl Keywords}: Bayesian nonparametrics, extremes, shifted gamma gamma distribution, non conjugate analysis, adaptive MCMC.

\section{Introduction}
\label{sec:intro}

Heavy or fat-tailed data have been studied in different fields, such as finance \citep{mandelbrot&hudson:08}, in the study of prices and returns, and insurance \citep{latchman&al:08}, in the study of claims from natural disasters such as earthquakes and floods. 

According to \cite{cooke&nieboer:11}, there are three main ways of defining a heavy tailed distribution. The first one is based on the kurtosis coefficient; leptokurtic distributions have fatter tails than normal distributions, so a distribution with kurtosis coefficient larger than zero is said to be heavy tail. The second definition is based on the theory of regular variation. This characterises the rate at which probabilities in the tail go to zero, that is, if $\lim_{x\to\infty}\P(X>x)/x^{-\alpha}\to 1$ then the probability distribution is said to be of regular variation with tail index $\alpha$; in particular, the moments of order $k\geq\alpha$ are all infinite. 
The third definition is based on subexponentiality, that is, the tail of the sum of independent variables behaves like the tail of the maximum. In notation, $\P(X_1+\cdots+X_n>x)\approx\P(\max\{X_1,\ldots,X_n\}>x)$. In general, the set of regular variation distributions is a strict subclass of the subexponential distributions. 

Estimating the tail of a distribution using extreme value theory goes back to \cite{davison&smith:90} who proposed methods for exceedances over thresholds and \cite{mcneil:97} who used the generalised pareto distribution (GPD) in the context of insurance data. To avoid wasting valuable information in the body of the distribution (below the threshold), mixture distributions have been proposed to model the entire set of data. For example, \cite{behrens&al:04} and \cite{nascimento&al:12} considered gamma and mixtures of gammas, respectively, combined with a GPD and used a bayesian approach for parameter estimation including the threshold; \cite{carreau&bengio:09} used a normal combined with a GPD and called their model hybrid pareto; and \cite{frigessi&al:02} proposed a dynamic mixture between a weibull and a GPD. In the discrete world, \cite{lee&eastoe:20} considered a mixture of geometric and an integer version of the GPD. On the other hand, \cite{bladt&rojas:08} suggested a scale mixture of phase-type distributions. 

Recently, two proposals have been introduced in the context of bayesian nonparametric (BNP) models. \cite{ayala&al:22} represented a phase-type distribution as an infinite mixture and connected the result to the dirichlet process mixture model to simplify bayesian inference. Finally, \cite{palacios&al:25} characterised the tails of the  normalised generalised gamma \citep{lijoi&al:07} mixture models and proposed scale and shape parameter mixtures of pareto type kernels. These authors found that if the kernel is heavy tail (regular variant), the BNP shape mixtures, regardless of the mixing distribution, are also heavy tail. 

For the purpose of this article we will refer to a heavy tail distribution if it belongs to the class of regular variation. In particular, we propose a BNP mixture model based on shifted gamma-gamma (SGG) densities with location, shape, tail, and scale parameters, and use the (two parameters) Poisson-Dirichlet (PD) process as mixing distribution for the four parameters vector. Exploiting the discreteness (clustering) properties of the PD process, the idea is to characterise the data with as few as possible mixture components, where some of them will be associated to model the bulk and others associated to the tail of the data, by means of the tail parameter of the SGG which controls the heaviness of the tails. At the same time, threshold parameters are not required to be fixed a-priori since these can be accounted for the location parameters of the SGG densities, which will also be estimated by the model. The main challenge of the proposal is to construct an efficient posterior sampler since the SGG does not admit conjugate priors for the four parameters vector. 

The contents of the rest of the paper is as follows: Section \ref{sec:heavy} introduces the SGG density. In Section \ref{sec:mixture} we present the bayesian nonparametric SGG mixture. In Section \ref{sec:posterior} we characterise the posterior distribution and propose an adaptive MCMC sampler. In section \ref{sec:numerical} we illustrate the performance of our model with simulated and real datasets. We finally conclude in Section \ref{sec:concl}.

\section{Heavy tail densities}
\label{sec:heavy}

Let $X$ be a continuous random variable with GPD density given by \citep[e.g.][]{johnson&al:95}
\begin{equation}
\label{eq:gpd}
f(x\mid\mu,\sigma,\xi)=\frac{1}{\sigma}\left\{1+\frac{\xi(x-\mu)}{\sigma}\right\}^{-\left(1+1/\xi\right)}I_{[\mu,\infty)}(x),
\end{equation}
where $\mu\in\RB$ is the location parameter, $\sigma>0$ is the scale parameter and $\xi\geq 0$ is the shape/tail parameter that has been constrained to nonnegative values so that the density has no bounded upper support. When $\xi=0$ the density \eqref{eq:gpd} reduces to a shifted exponential (light tail) $f(x\mid\mu,\sigma)=\exp\{-(x-\mu)/\sigma\}/\sigma$. For $\xi<0$ the GPD density \eqref{eq:gpd} remains valid, but the support is constrained to the interval $[\mu,\mu-\sigma/\xi]$ (short tail). 

The density \eqref{eq:gpd} belongs to the regular variation class if $\xi>0$, that is, in this case it is considered to have a heavy tail. For smaller values of $\xi$ the tail becomes less heavy, in fact, for $\xi<1$ the first moment exists taking the value $\E(X)=\mu+\sigma/(1-\xi)$, for the second moment to exist we need a smaller value $\xi<1/2$, in which case $\V(X)=\sigma^2/\{(1-\xi)^2(1-2\xi)\}$. For the third and fourth moments to exist, we need $\xi<1/3$ and $\xi<1/4$, respectively, and so on. In the limit, when $\xi\to 0$, the moment generating function exists, which means that all moments exist and the model has a light tail. In summary, the heaviness of the tail is controlled by the shape/tail parameter $\xi$. 

The GPD density is closely related to a gamma-gamma ($\gga$) density. This is defined \cite[e.g.][]{nieto:25} as follows. Let $X\mid Y\sim\ga(\gamma,y)$ and $Y\sim\ga(\alpha,\beta)$, where the gamma density is parametrised in terms of shape and rate, that is, $\E(Y)=\alpha/\beta$. Then marginally, $X\sim\gga(\gamma,\alpha,\beta)$ with nonnegative parameters $\gamma,\alpha$ and $\beta$. In practice $\gamma$ is associated to a sample size and therefore usually takes positive integer values, but the model is well defined for nonnegative values. The GG density is given by 
\begin{equation}
\label{eq:gga}
f(x\mid \gamma,\alpha,\beta)=\frac{\beta^\alpha}{\Gamma(\alpha)}\frac{\Gamma(\alpha+\gamma)}{\Gamma(\gamma)}\frac{x^{\gamma-1}}{(\beta+x)^{\alpha+\gamma}}I_{[0,\infty)}(x).
\end{equation}
The GG density is of regular variation for all $\alpha>0$ and has no moments if $\alpha\leq 1$. For $\alpha>1$ the tail becomes less heavy. The first moment $\E(X)=\beta\gamma/(\alpha-1)$ exists for $\alpha>1$, and the second moment $\V(X)=\beta^2\gamma(\gamma+\alpha-1)/\{(\alpha-1)^2(\alpha-2)\}$ exists for $\alpha>2$. In general, the $k^{th}$ moment exists for $\alpha>k$. 

It is not difficult to prove that a $\gga(\gamma,\alpha,\beta)$ with $\gamma=1$ simplifies to a GPD with parameters $\mu=0$, $\sigma=\beta/\alpha$ and $\xi=1/\alpha$. To make a complete connection between models \eqref{eq:gpd} and \eqref{eq:gga} we require a location parameter $\mu\in\RB$ that shifts the GG density. For that we take $X-\mu\mid Y\sim\ga(\gamma,y)$ and $Y\sim\ga(\alpha,\beta)$. In this case, the marginal density for $X$ is
\begin{equation}
\label{eq:sgg}
f(x\mid\mu,\gamma,\alpha,\beta)=\frac{\beta^\alpha}{\Gamma(\alpha)}\frac{\Gamma(\alpha+\gamma)}{\Gamma(\gamma)}\frac{(x-\mu)^{\gamma-1}}{(\beta+x-\mu)^{\alpha+\gamma}}I_{[\mu,\infty)}(x).
%f(x\mid\mu,\gamma,\alpha,\beta)=\frac{\beta^\alpha\Gamma(\alpha+\gamma)(x-\mu)^{\gamma-1}}{\Gamma(\alpha)\Gamma(\gamma)(\beta+x-\mu)^{\alpha+\gamma}}I_{[\mu,\infty)}(x).
\end{equation}
We will refer to this density as shifted gamma-gamma and denote it as $\sgg(\mu,\gamma,\alpha,\beta)$, where $\mu$ is the location parameter, $\gamma$ is the shape parameter, $\alpha$ is the tail parameter and $\beta$ the scale parameter. In this case, the SGG model \eqref{eq:sgg} generalises both the GPD model \eqref{eq:gpd} and the GG model \eqref{eq:gga}. Specifically, $\sgg(0,\gamma,\alpha,\beta)\equiv\gga(\gamma,\alpha,\beta)$ and $\sgg(\mu,1,\alpha,\beta)\equiv\gpd(\mu,\beta/\alpha,1/\alpha)$.  

\begin{figure}
\centerline{\includegraphics[scale=0.45]{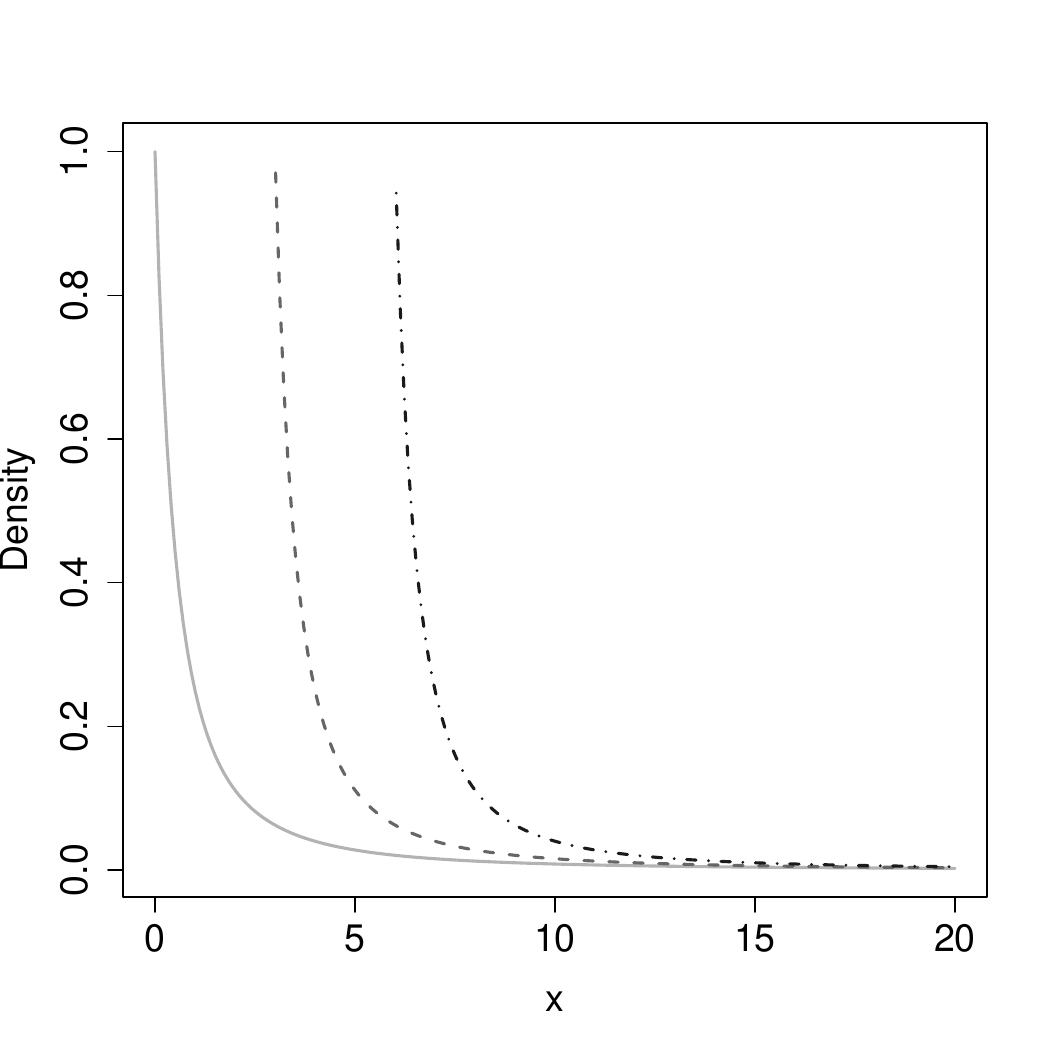}
\includegraphics[scale=0.45]{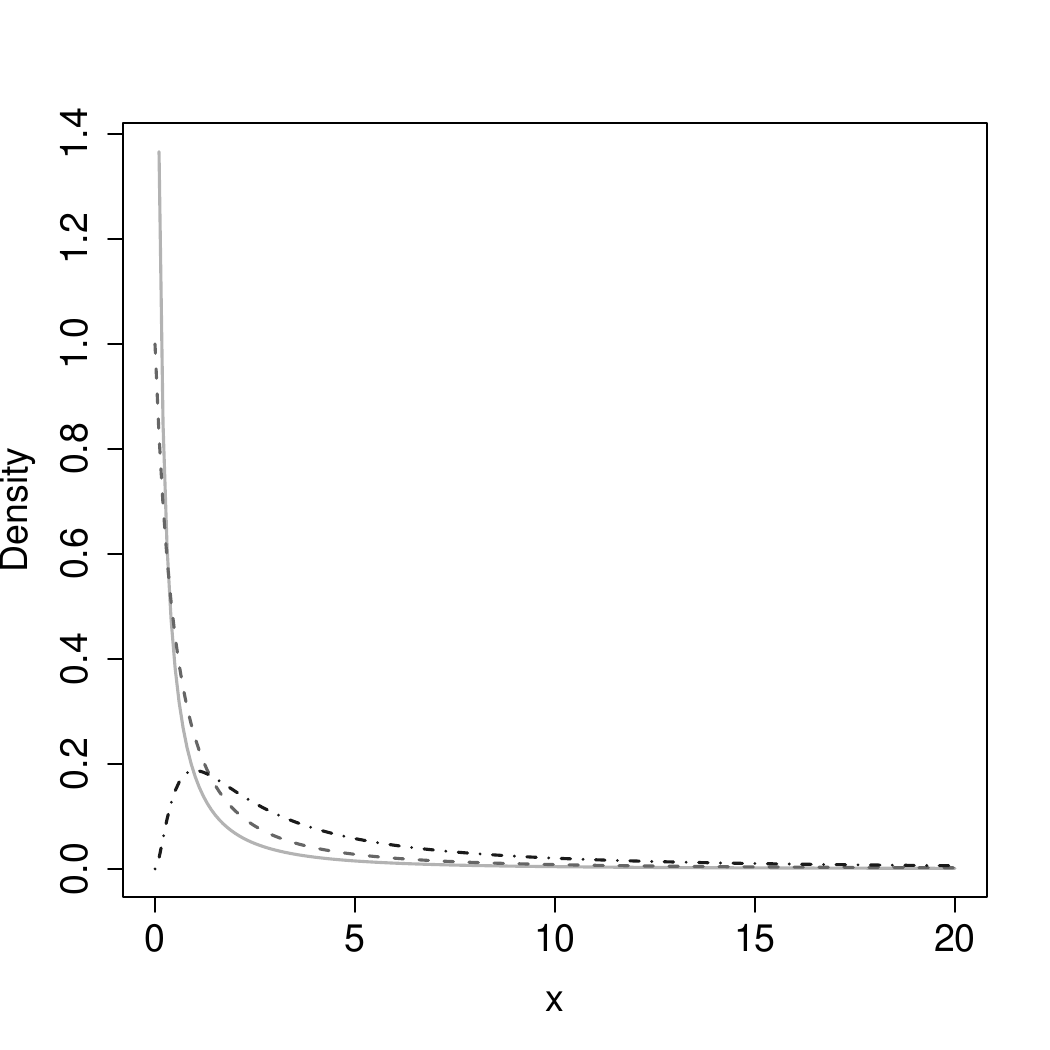}}
\centerline{\includegraphics[scale=0.45]{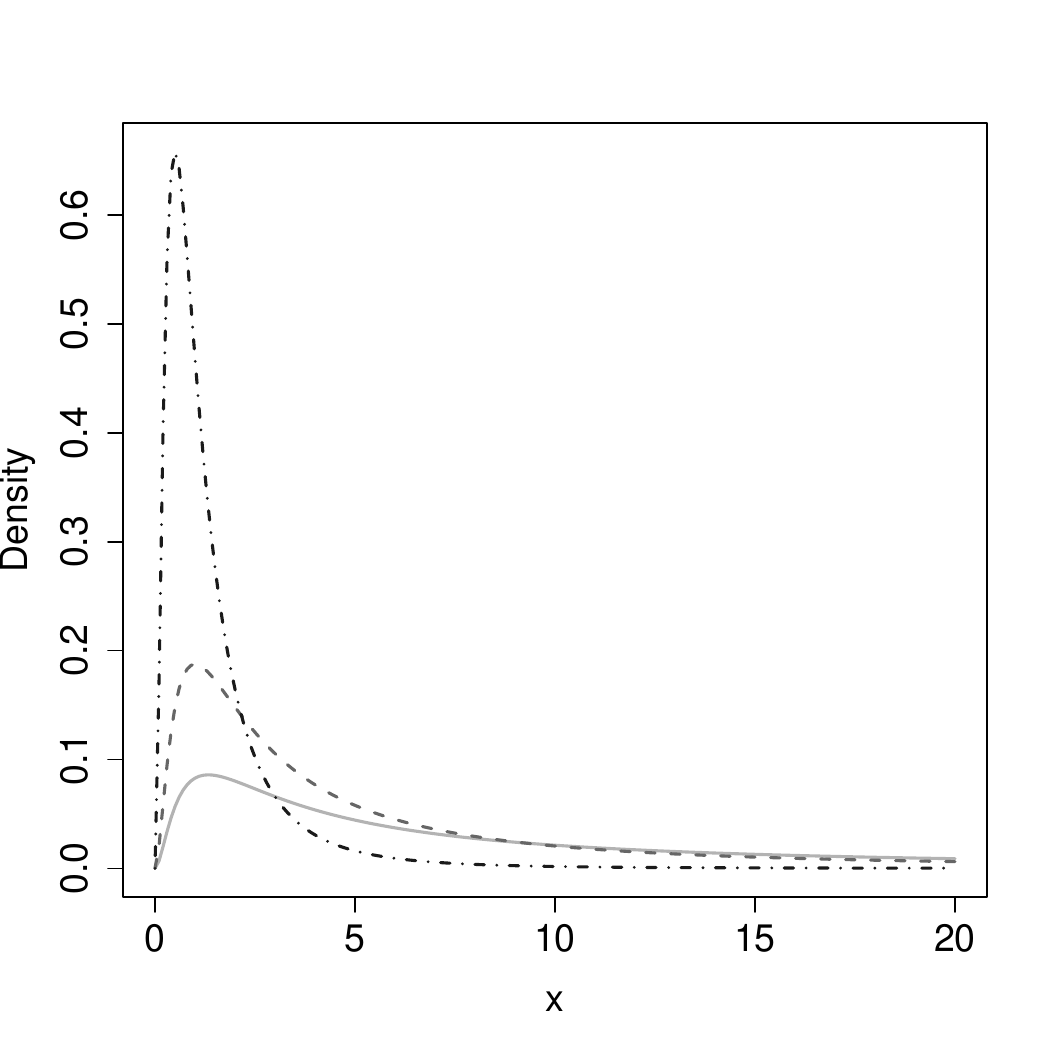}
\includegraphics[scale=0.45]{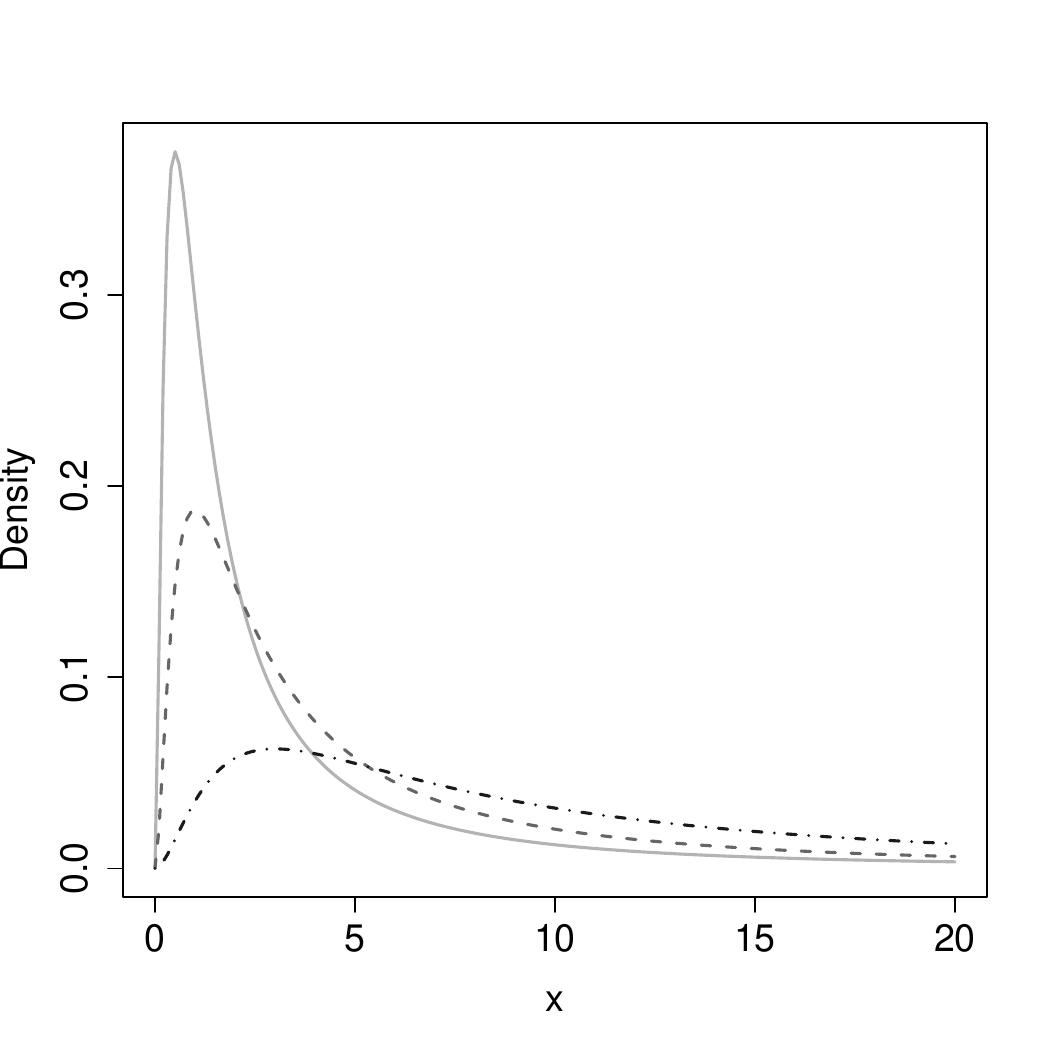}}
\caption{{\small SGG densities for varying parameters. $\mu\in\{0,3,6\}$ (top left); $\gamma\in\{0.5,1,3\}$ (top right); $\alpha\in\{0.5,1.3\}$ (bottom left); $\beta\in\{0.5,1,3\}$ (bottom right). Darker colours mean larger parameter values.}}
\label{fig:sgg}
\end{figure}

The role played by the different parameters $(\mu,\gamma,\alpha,\beta)$, in the $\sgg$ model \eqref{eq:sgg}, is shown in Figure \ref{fig:sgg}. In the top left panel we take varying $\mu\in\{0,3,6\}$ and fixed $\gamma=\alpha=\beta=1$; here $\mu$ is a location parameter that shifts the whole density to the right for larger values. In the top right panel, we take varying $\gamma\in\{0.5,1,3\}$ and fixed $\mu=0$, $\alpha=\beta=1$; here $\gamma$ is a shape parameter, for $\gamma<1$ the density has a vertical asymptote at zero, for $\gamma=1$ the density takes the value of one at zero, and for $\gamma>1$ the density pushes the mass away from zero inducing a positive mode. In the bottom left panel, we take varying $\alpha\in\{0.5,1,3\}$ and fixed $\mu=0$, $\gamma=3$ and $\beta=1$; here $\alpha$ is a tail parameter that induces a less heavy tail for $\alpha>1$ and a heavier tail for $\alpha\leq 1$. Finally, in the bottom right panel, we take varying $\beta\in\{0.5,1,3\}$ and fixed $\mu=0$, $\gamma=3$ and $\alpha=1$; here $\beta$ is a scale parameter where smaller/larger values produce a faster/slower decay.

\section{Mixture model}
\label{sec:mixture}

There are different classes of bayesian noparametric priors, but most of them are almost surely discrete. This feature, combined in a mixture model, has recently been exploited for modelling purposes because it can be used to identify underlying groups in the data \citep[e.g.][]{nieto&contreras:14}. On the other hand, recent studies made by \cite{palacios&al:25} show that BNP mixtures have heavy tails as long as the mixing kernel has heavy tails. 

In general, an almost surely discrete random probability measure $G$ can be written as $G(\cdot)=\sum_{j=1}^\infty W_j\delta_{Z_j}(\cdot)$, where $\{W_j\}$ is a sequence of nonnegative random variables such that $\sum_{j=1}^\infty W_j=1$ a.s., $\{Z_j\}$ is a sequence of random variables taking values in $\RB$, and $\delta_{Z}(\cdot)$ is the dirac measure. The Poisson-Dirichlet process is a rich class that includes the Dirichlet process \citep{ferguson:73} and the normalised stable process \citep{kingman:75}. 

The PD process admits a constructive representations as a stick breaking prior \citep[e.g.][]{ishwaran&james:01}. $G$ is said to be a PD process with parameters $\nu\in(0,1)$, $\kappa>-\nu$ and centering measure $G_0=\E(G)$, denoted as $G\sim\PD(\nu,\kappa,G_0)$, if for $j=1,2,\ldots$ the random locations are $Z_j\simiid G_0$, and the random weights are $W_j=V_j\prod_{k=1}^{j-1}(1-V_k)$ with $V_j\simiid\be(1-\nu,\kappa+j\nu)$, where $\be(\alpha,\beta)$ denotes a beta distribution with mean $\alpha/(\alpha+\beta)$. When $\nu=0$ the PD process reduces to a Dirichlet process (DP), whereas for $\kappa=0$ the PD becomes a normalised stable (NS) process. 

If we have realisations from a PD process, say $\theta_i\mid G\simiid G$ for $i=1,\ldots,n$ and $G\sim\PD(\nu,\kappa,G_0)$, then marginally $\theta_i\sim G_0$, and the joint (predictive) distribution of $\btheta=\{\theta_i\}$ can be represented as a generalised P\'olya urn of the following form \citep{pitman:95}
\begin{equation}
\label{eq:purn}
f(\theta_i\mid\btheta_{-i})=\frac{\kappa+\nu m_i}{\kappa+n-1}g_0(\theta_i)+\sum_{j=1}^{m_i}\frac{n_{j,i}^*-\nu}{\kappa+n-1}\delta_{\theta_{j,i}^*}(\theta_i),
\end{equation}
for $i=1,\ldots,n$, where $\btheta_{-i}$ denotes the vector $\btheta$ that excludes $\theta_i$, $(\theta_{1,i}^*,\ldots,\theta_{m_i,i}^*)$ denote the unique values in $\btheta_{-i}$, each occurring with frequency $n_{j,i}^*$ for $j=1,\ldots,m_i$, which satisfy the condition $\sum_{j=1}^{m_i}n_{j,i}^*=n-1$, and $g_0$ is the density associated with $G_0$. The number of unique values $m$ in $\btheta$ is controlled by both $\nu$ and $\kappa$, where larger values of $\nu$ and/or $\kappa$ produces larger $m$. 

We are now in a position to write our mixture model for the analysis of heavy tail data. Let $\theta_i=(\mu_i,\gamma_i,\alpha_i,\beta_i)$ be a four dimensional parameter vector, then we assume 
\begin{align}
\nonumber
X_i\mid\theta_i&\simind\sgg(\mu_i,\gamma_i,\alpha_i,\beta_i)\\
\label{eq:mixm}
\theta_i\mid G &\simiid G,\quad i=1,\ldots,n\\
\nonumber
G &\sim\PD(\nu,\kappa,G_0),
\end{align}
where 
\begin{equation}
\label{eq:g0}
g_0(\theta)=\no(\mu\mid a_{\mu},b_{\mu})\ga(\gamma\mid a_{\gamma},b_{\gamma})\ga(\alpha\mid a_{\alpha},b_{\alpha})\ga(\beta\mid a_{\beta},b_{\beta}), 
\end{equation}
where $\no(a,b)$ denotes a normal distribution with mean $a$ and variance $b$. 

Since the PD prior induces ties in the $\theta_i$'s, the number of mixture components $m$, in the mixture, depends on $(\nu,\kappa)$ and the hyper parameters $a_{\eta},b_{\eta}$ for $\eta\in\{\mu,\gamma,\alpha,\beta\}$. The idea is to choose these hyper parameters so that we have few mixture components where some of them are associated to model the bulk with less heavy tails (small $\mu_j$ and $\alpha_j>1$) and with few others the heavier tails (large $\mu_j$ and $\alpha_j\leq 1$) of the data. 
%In the presence of heavier tails, the location parameter $\mu_j$ will determine the value of the support from where a data point is considered extreme. 
%If the data do not admit heavier tails, all $\alpha_j$ for $j=1,\ldots,m$ will be greater than one. 
The shape and scale parameters $\gamma_j$ and $\beta_j$ are of lesser importance but allow for greater flexibility in the model.

\section{Posterior characterisation}
\label{sec:posterior}

There are different ways of characterising the posterior distribution of model \eqref{eq:mixm}. Via a conditional sampler that is based on obtaining the posterior law of the process $G$, as in \cite{barrios&al:13}, and via the marginal sampler which is based on marginalising the stochastic process $G$, as in \cite{favaro&teh:13}. Here we propose the latter, since our interest is to characterise the clustering structure induced by the model and to potentially identify those groups associated to model either the bulk or the heavier tail of the data. 

The likelihood of the model is defined by the first equation in \eqref{eq:mixm}. However, as also noted in Section \ref{sec:heavy}, the SGG model can be obtained as the marginal distribution for $X_i$ of the joint for $(X_i,Y_i)$ given by $X_i-\mu_i\mid \mu_i,\gamma_i,Y_i\simind\ga(\gamma_i,y_i)$ and $Y_i\mid\alpha_i,\beta_i\simind\ga(\alpha_i,\beta_i)$ for $i=1,\ldots,n$. Therefore, the posterior becomes simpler if we consider the augmented likelihood for $(\bX,\bY)\mid\btheta$ given by
%$$f(\bx,\by\mid\btheta)=\prod_{i=1}^n\ga(x_i-\mu_i\mid\gamma_i,y_i)\ga(y_i\mid\alpha_i,\beta_i).$$
\begin{equation}
\label{eq:lik}
%X_i-\mu_i\mid \mu_i,\gamma_i,Y_i\simind\ga(\gamma_i,y_i)\quad\mbox{and}\quad Y_i\mid\alpha_i,\beta_i\simind\ga(\alpha_i,\beta_i).
f(\bx,\by\mid\btheta)=\prod_{i=1}^n f(x_i,y_i\mid\theta_i)=\prod_{i=1}^n\ga(x_i-\mu_i\mid\gamma_i,y_i)\ga(y_i\mid\alpha_i,\beta_i). 
\end{equation}

The posterior distribution is obtained proportionally by the product of \eqref{eq:lik} and \eqref{eq:purn}. Therefore, the posterior conditional distributions for each $\theta_i$ are given by
$$f(\theta_i\mid\bx,\by,\btheta_{-i})\propto (\kappa+\nu\, m_i) g_0(\theta_i)f(x_i,y_i\mid\theta_i)+\sum_{j=1}^{m_i}(n_{j,i}^*-\nu)\delta_{\theta_{j,i}^*}(\theta_i)f(x_i,y_i\mid\theta_i),$$
for $i=1,\ldots,n$. Since $g_0$ is clearly not conjugate with respect to the likelihood, to sample from these posterior conditional distributions, we resource to the generalisation of \cite{neal:00}'s Algorithm 8, also suggested by \cite{favaro&teh:13}. Initialise the algorithm by sampling $\theta_i$ values, for $i=1,\ldots,n$, from the marginal prior $g_0$, in \eqref{eq:g0}. Then the algorithm proceeds as follows: 
\begin{enumerate}
\item[(i)] For each $X_i$, $i=1,\ldots,n$, sample a latent value $Y_i$ from its conditional distribution 
$$f(y_i\mid x_i,\theta_i)=\ga(y_i\mid\gamma_i+\alpha_i,x_i-\mu_i+\beta_i).$$
\item[(ii)] For each $i=1,\ldots,n$, sample $r$ auxiliary values $\btheta^{\star}=\{\theta_{m_i+1}^{\star},\ldots,\theta_{m_i+r}^{\star}\}$ from $g_0$, given in \eqref{eq:g0}. Values not selected in step (iii) can be reused. 
\item[(iii)] Draw $\theta_i$, $i=1,\ldots,n$, from 
$$\hspace{-11.5cm}f(\theta_i\mid\bx,\by,\btheta_{-i},\btheta^{\star})=$$ 
$$\hspace{1cm}\frac{1}{c_i}\left[\sum_{j=1}^{m_i}(n_{j,i}^*-\nu) f(x_i,y_i\mid\theta_{j,i}^*)\delta_{\theta_{j,i}^*}(\theta_i)+\sum_{j=m_i+1}^{m_i+r}\left(\frac{\kappa+\nu m_i}{r}\right)f(x_i,y_i\mid\theta_{j}^{\star})\delta_{\theta_{j}^{\star}}(\theta_i)\right],$$
where $c_i=\sum_{j=1}^{m_i}(n_{j,i}^*-\nu)f(x_i,y_i\mid\theta_{j,i}^*)+\sum_{j=m_i+1}^{m_i+r} \{(\kappa+\nu m_i)/r\}f(x_i,y_i\mid\theta_{j}^{\star})$. 
\item[(iv)] Compute the unique values $(\theta_1^*,\ldots,\theta_m^*)$ in $\btheta$ and re-sample each $\theta_j^*$, $j=1,\ldots,m$ from 
$$f(\theta_j^*\mid \data)\propto g_0(\theta_j^*)\prod_{\{i:\theta_i=\theta_j^*\}}f(x_i,y_i\mid\theta_j^*),$$
where $\data$ are the data points assigned to the same cluster.  
\end{enumerate}

The number of clusters in the nonparametric mixture \eqref{eq:mixm} is determined by the PD prior. In particular, small/large values of $\nu,\kappa$ produce small/large number of clusters. Instead of fixing them a-priori, we assign a joint hyper-prior of the form $f(\nu,\kappa)=\be(\nu\mid a_{\nu},b_{\nu})\ga(\kappa+\nu\mid a_{\kappa},b_{\kappa})$. This prior is updated by the exchangeable partition probability function induced by the PD process, which is given by 
$$f(n_1^*,\ldots,n_m^*\mid\nu)=\frac{\Gamma(\kappa+1)}{\Gamma(\kappa+n)}\left\{\prod_{j=1}^{m-1}(\kappa+j\nu)\right\}\left\{\prod_{j=1}^m\frac{\Gamma(n_j^*-\nu)}{\Gamma(1-\nu)}\right\}.$$ Therefore, we add two further steps to the sampling algorithm as follows.
\begin{enumerate}
\item[(v)] Sample a new value of $\nu$ from 
$$\hspace{-5cm}f(\nu\mid\kappa,\data)\propto \nu^{a_{\nu}-1}(1-\nu)^{b_{\nu}-1}(\kappa+\nu)^{a_{\kappa}-1}e^{-b_{\kappa}\nu}$$
$$\hspace{7mm}\left\{\prod_{j=1}^{m-1}(\kappa+j\nu)\right\}\left\{\prod_{j=1}^m\frac{\Gamma(n_j^*-\nu)}{\Gamma(1-\nu)}\right\}I_{(\max\{-\kappa,0\},1)}(\nu).$$
\item[(vi)] Sample a new value of $\kappa$ from 
$$f(\kappa\mid\nu,\data)\propto (\kappa+\nu)^{a_{\kappa}-1}e^{-b_{\kappa}\kappa}\,\frac{\Gamma(\kappa+1)}{\Gamma(\kappa+n)}\left\{\prod_{j=1}^{m-1}(\kappa+j\nu)\right\}I_{(-\nu,\infty)}(\kappa)$$
\end{enumerate}

Performing steps (i)--(iii) is straightforward, however, sampling from (iv)--(vi) require Metropolis-Hastings steps \citep[e.g.][]{robert&casella:10}, due to the nonconjugacy between the sampling distribution and the centering measure. 
We suggest using a random walk proposal distribution for each element of $\theta_j^*$, for $\nu$ and for $\kappa$. Let $\vartheta\in\{\mu_j^*,\gamma_j^*,\alpha_j^*,\beta_j^*,\nu,\kappa\}$. At iteration $(t+1)$ draw a proposal $\vartheta$ from $\un(\max\{\vartheta^{(t)}-\delta_{\vartheta},L_{\vartheta}\},\min\{\vartheta^{*(t)}+\delta_{\vartheta},U_{\vartheta}\})$ and accept it with probability $\min\{1,f(\vartheta\mid \data)/f(\vartheta^{(t)}\mid \data)\}$. Here, $L_{\vartheta}$ is a lower bound that takes the values $L_{\vartheta}=-\infty$ for $\vartheta_j=\mu_j^*$,  $L_{\vartheta}=\max\{-\kappa,0\}$ for $\vartheta=\nu$, $L_{\vartheta}=-\nu$ for $\vartheta=\kappa$ and $\vartheta=0$ for the other three parameters; $U_{\vartheta}$ is an upper bound that takes the values $U_{\vartheta}=\min\{x_i:\theta_i=\theta_j^*\}$ for $\vartheta_j=\mu_j^*$, $U_{\vartheta}=1$ for $\vartheta_j=\nu$, and $U_{\vartheta}=\infty$ for the other four parameters. 

Parameters $\delta_\vartheta$ are tuning parameters that control acceptance probability. Instead of fixing them, we suggest adapting them every certain number of iterations to achieve a target acceptance rate. We consider as target interval $[0.3,0.4]$ which, according to \cite{robert&casella:10}, defines optimal acceptance rates. Specifically, we use batches of 80 iterations and for each batch $b$, we compute the acceptance rate $AR_{\vartheta}^{(b)}$ and decrease $\delta_{\vartheta}^{(b+1)}=\delta_{\vartheta}^{(b)}2^{-1/\sqrt{b}}$ if $AR_{\vartheta}^{(b)}<0.3$ and increase $\delta_{\vartheta}^{(b+1)}=\delta_{\vartheta}^{(b)}2^{1/\sqrt{b}}$ if $AR_{\vartheta}^{(b)}>0.4$. We use $\delta_{\vartheta}^{(1)}=1$ as the starting value.
Our adaptation proposal is based on a diminishing adaptation condition \citep{roberts&rosenthal:09} such that when the batch $b\to\infty$ then $2^{1/\sqrt{b}}\to 1$ and $2^{-1/\sqrt{b}}\to 1$, which implies that eventually $\delta_{\vartheta}^{(b)}$ will stop moving. 

Note that the number of parameters $m$ for $\mu_j^*$, $\gamma_j^*$, $\alpha_j^*$, $\beta_j^*$, $j=1,\ldots,m$, varies between iterations. For these, we use a single tuning parameter $\delta_\vartheta$ and the acceptance rate $AR_{\vartheta}^{(b)}$ is the average of the acceptance rates in batch $b$ for all $j=1,\ldots,m$.

\section{Numerical analyses}
\label{sec:numerical}

\subsection{Simulation Study}

The objective of this study is two fold. First, test the posterior MCMC algorithm outlined in Section \ref{sec:posterior} and second, prove that under a control scenario, the model is able to determine the bulk and tail of the data appropriately. 

\begin{figure}
\centerline{\includegraphics[scale=0.9]{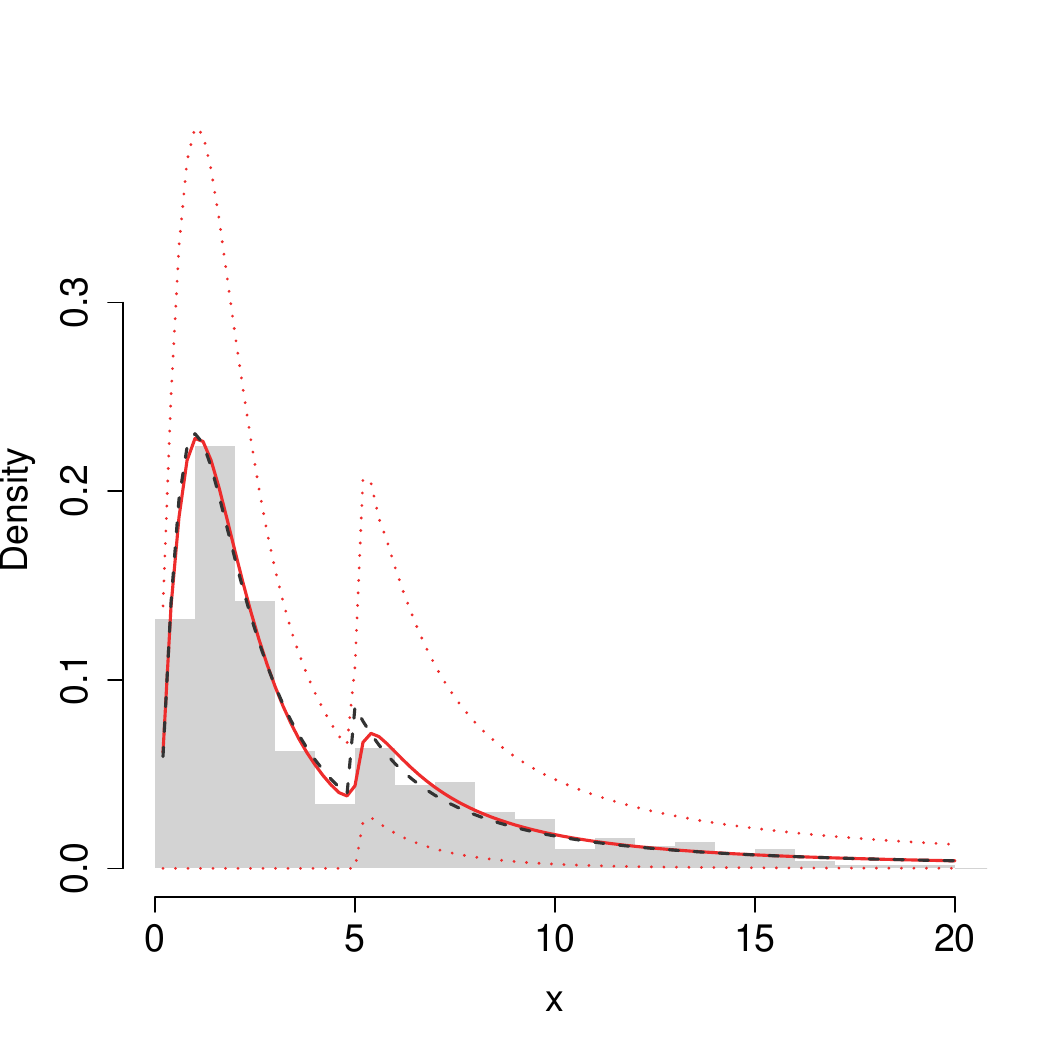}}
\caption{{\small Simulated data and fit. Histogram of data (background), true density (dashed line), posterior density estimate (solid line), 95\% posterior CI (dotted lines).}}
\label{fig:dens}
\end{figure}

For that, let us consider a two component mixture of SGG densities of the form $$f(x)=\pi\sgg(x\mid\mu_1,\gamma_1,\alpha_1,\beta_1)+(1-\pi)\sgg(x\mid\mu_2,\gamma_2,\alpha_2,\beta_2),$$ where $\pi=0.7$, $\mu_1=0$, $\gamma_1=3$, $\alpha_1=3$, $\beta_1=2$, $\mu_2=5$, $\gamma_2=1$, $\alpha_2=0.5$, $\beta_2=3$. The resulting density is presented as the dashed line in Figure \ref{fig:dens}. It shows a bimodal behaviour with the first mode around one and the second mode at five. The first component of the mixture has a less heavy tail $(\alpha_1=3)$, where the first two moments exist, and the second component has a heavier tail $(\alpha_2=0.5)$. 

We took a sample of size $n=500$ and fitted our model with the following prior specifications: $a_{\mu}=0$, $b_{\mu}=100$, $a_{\gamma}=b_{\gamma}=1/2$, $a_{\alpha}=b_{\alpha}=1/2$ and $a_{\beta}=b_{\beta}=1/2$. For the parameters $(\nu,\kappa)$, of the PD process, we took $a_{\nu}=b_{\nu}=1/2$ and $a_{\kappa}=1$, $b_{\kappa}=2$ with the following combinations to compare: $\nu=0$ and prior on $\kappa$ to define a DP; $\kappa=0$ and prior on $\nu$ to define a NS process; and priors on both $\nu$ and $\kappa$ to define a PD process. 

The MCMC algorithm was run for $20,000$ iterations, a burn in of $5,000$ and a thinning of $5$, i.e., keeping one of every fifth iteration to compute posterior summaries. The running times are reported in the last column of Table \ref{tab:fits} and go from $3.69$ to $3.88$ minutes, which is quite fast. 
 
\begin{figure}
\centerline{\includegraphics[scale=0.45]{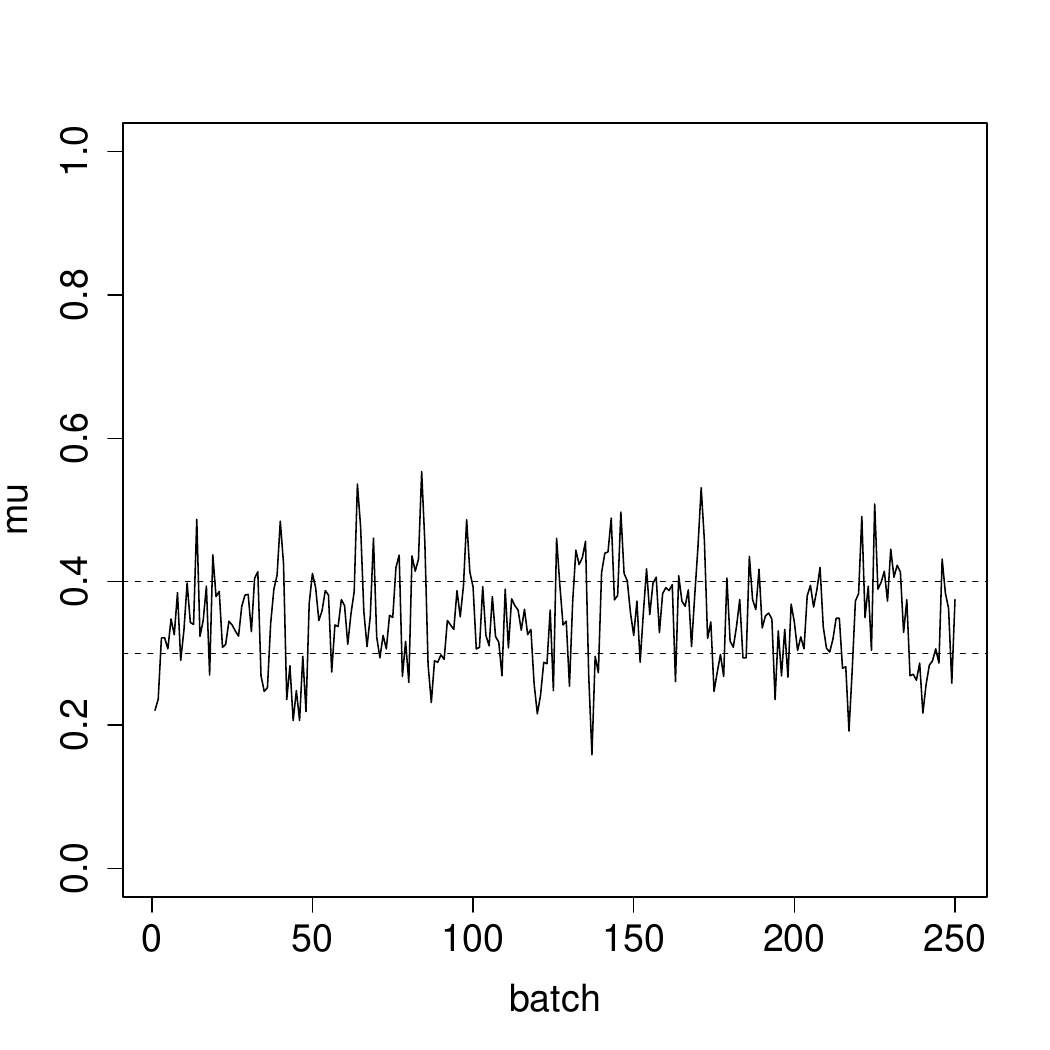}
\includegraphics[scale=0.45]{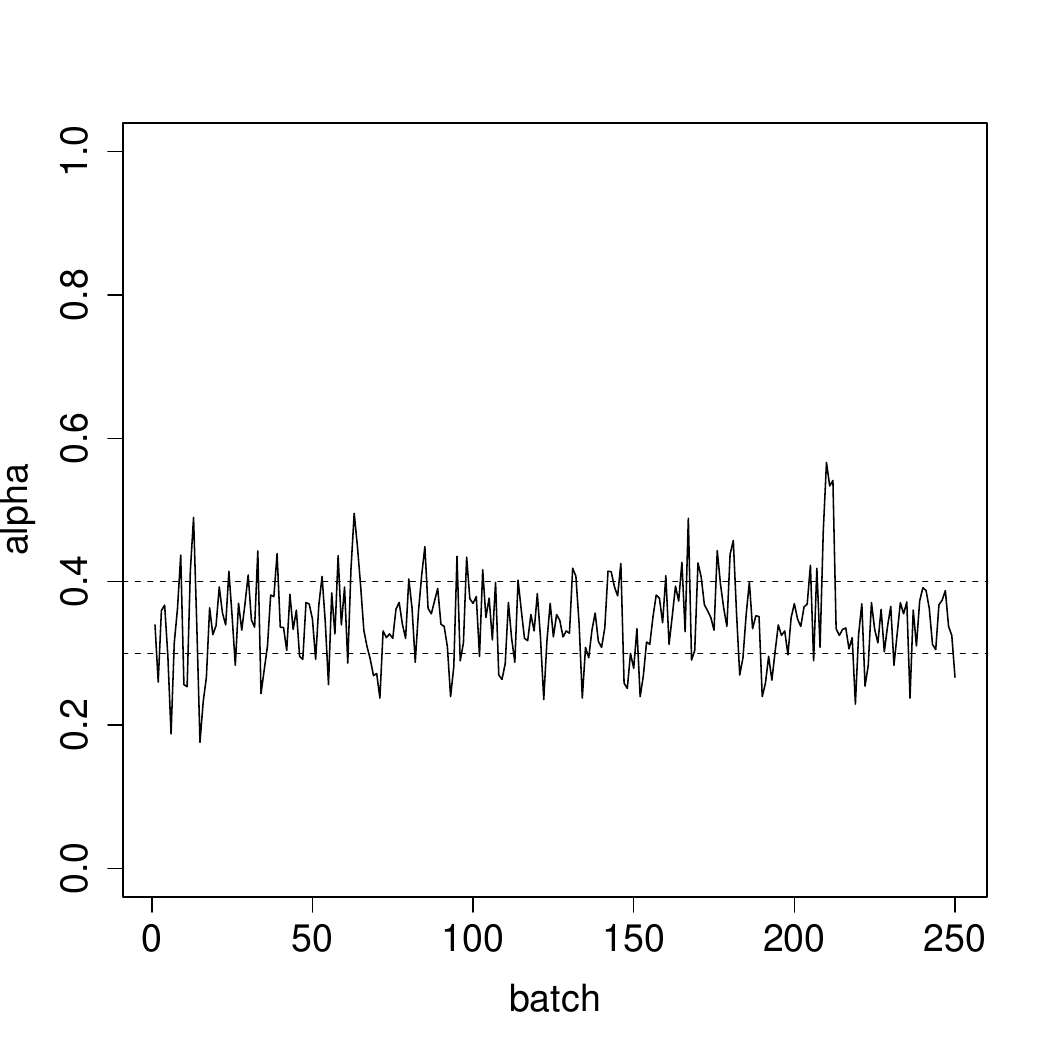}}
\caption{{\small Simulated data. Average acceptance rates in batches of size $b=50$ iterations for $\mu_j^*$ (left) and $\alpha_j^*$ (right), $j=1,\ldots,m$, when we take $\nu\sim\be(1/2,1/2)$. Target rate limits are shown as dotted horizontal lines.}}
\label{fig:ars}
\end{figure}

We assess the performance of our adapting MH tuning parameters $\delta_{\vartheta}$ by looking at the acceptance rates per batch, whose size was defined by $80$ iterations.  Figure \ref{fig:ars} shows the acceptance rates for parameters $\mu_j^*$ and $\alpha_j^*$. For both parameters, the acceptance rates oscillate around the target interval $[0.3,0.4]$, shown as horizontal dotted lines, as desired. 

Comparison among the different fittings is made using the logarithm of the pseudo marginal likelihood (LPML) proposed by \cite{geisser&eddy:79}. This is defined as $\mbox{LPML}=\sum_{i=1}^n\log(\mbox{CPO}_i)$, where $\mbox{CPO}_i=f(x_i\mid\bx_{-i})$. These CPO's can be easily approximated via Monte Carlo as follows. Given a posterior sample $\btheta^{(l)}$ and a sequence of latent variables $y_i^{(l)}$ for $l=1,\ldots,L$, then 
\begin{equation}
\label{eq:lpml}
\widehat{\mbox{CPO}}_i=\left(\frac{1}{L}\sum_{l=1}^L\frac{1}{f(x_i,y_i^{(l)}\mid\theta_i^{(l)})}\right)^{-1}.
\end{equation}
Additionally, we also computed the posterior expected values of $\mbox{AIC}=2m-2\log f(\bx\mid\btheta)$ and $\mbox{BIC}=m\log n-2\log f(\bx\mid\btheta)$, which penalise models with a larger number of parameters plus a large sample size, respectively. 

The fitting measures for the different specifications of the PD process are presented in Table \ref{tab:fits}. In the same table, we report the posterior mode of the number of groups in the BNP mixture as well as the running times. The worst fit is obtained by the NS specification, followed by the general PD, and finally the best fit is obtained by the  DP specification. In all cases the posterior mode for the number of groups is 2, with assigned probabilities to the mode 0.65, 0.56 and 0.69, respectively. 

\begin{figure}
\centerline{\includegraphics[scale=0.45]{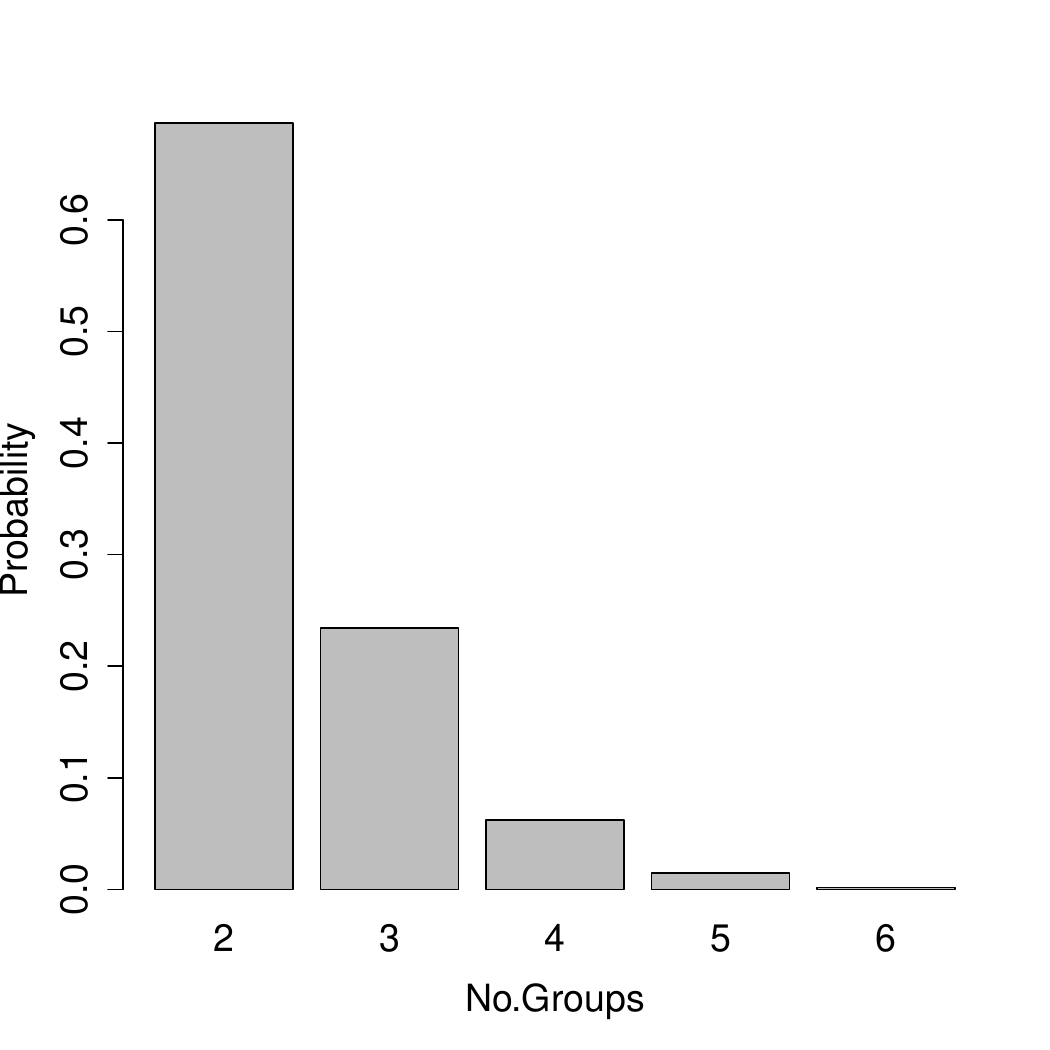}
\includegraphics[scale=0.45]{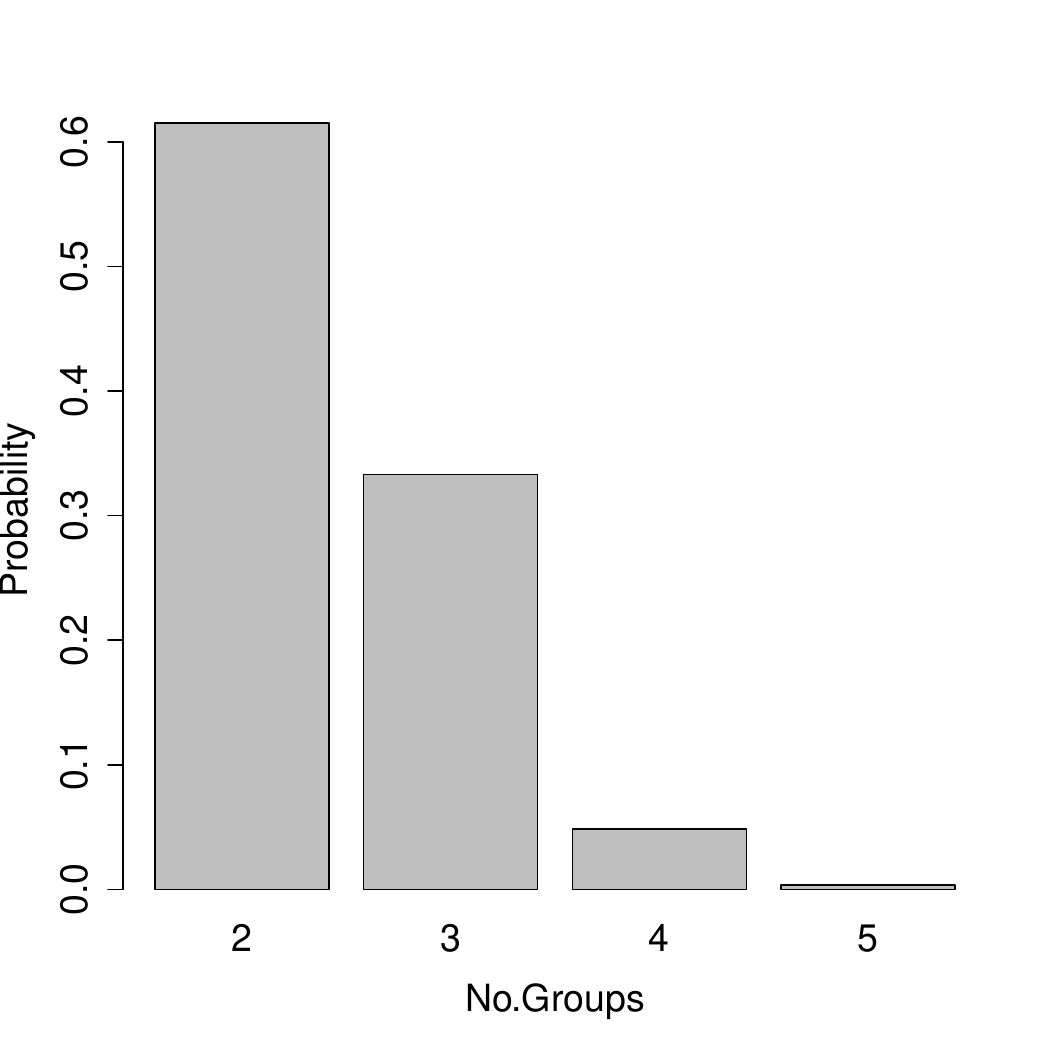}}
\caption{{\small Posterior distribution of the number of groups $m$ for the best fitting model. Left: simulated data, $\P(m=2)=0.69$; right: claims data, $\P(m=2)=0.62$.}}
\label{fig:n0}
\end{figure}

The posterior distribution of the number of groups, for the best fitting model (DP), is presented in the left panel in Figure \ref{fig:n0}. The posterior estimates of the DP precision parameter is $\widehat{\kappa}=0.32$ with $95\%$ CI $[0.04,0.91]$. 

\begin{figure}
\centerline{\includegraphics[scale=0.45]{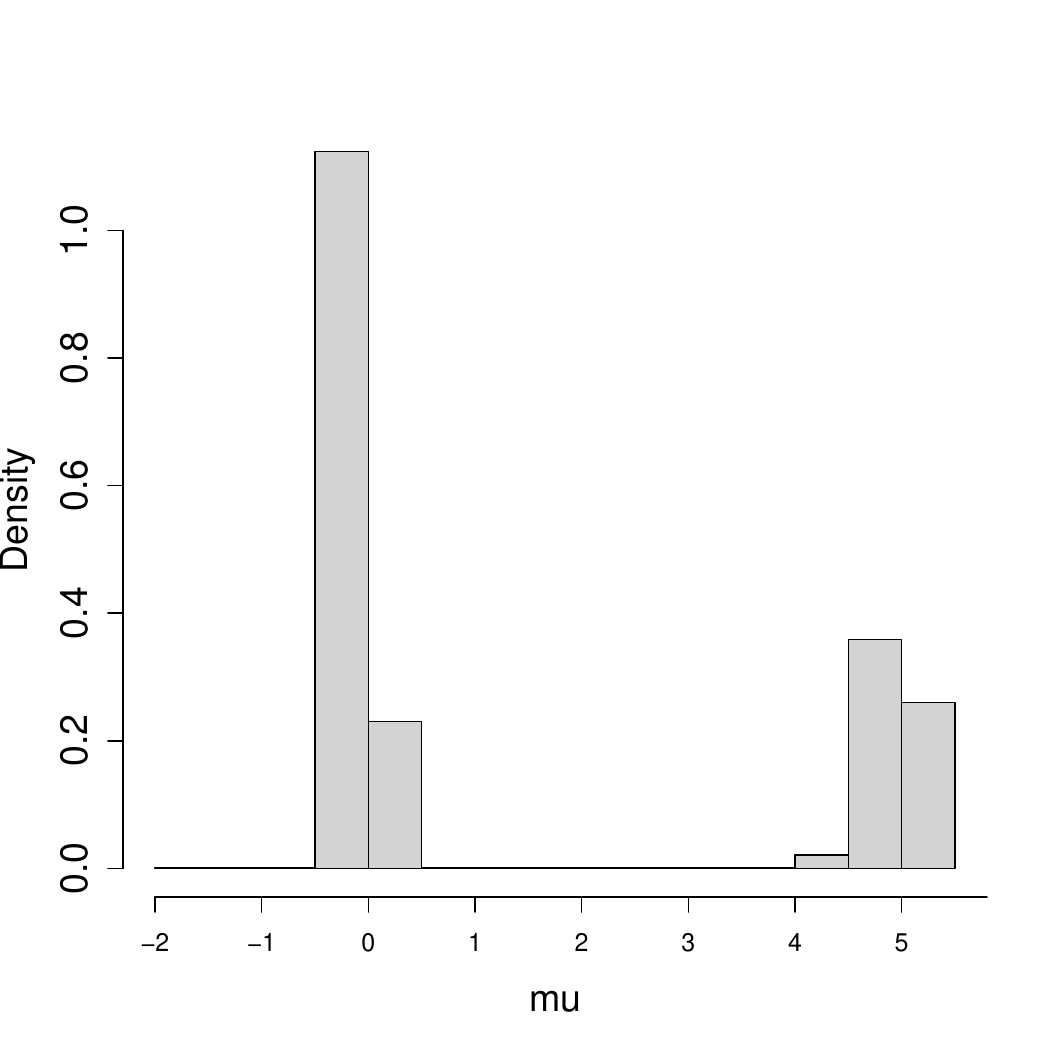}
\includegraphics[scale=0.45]{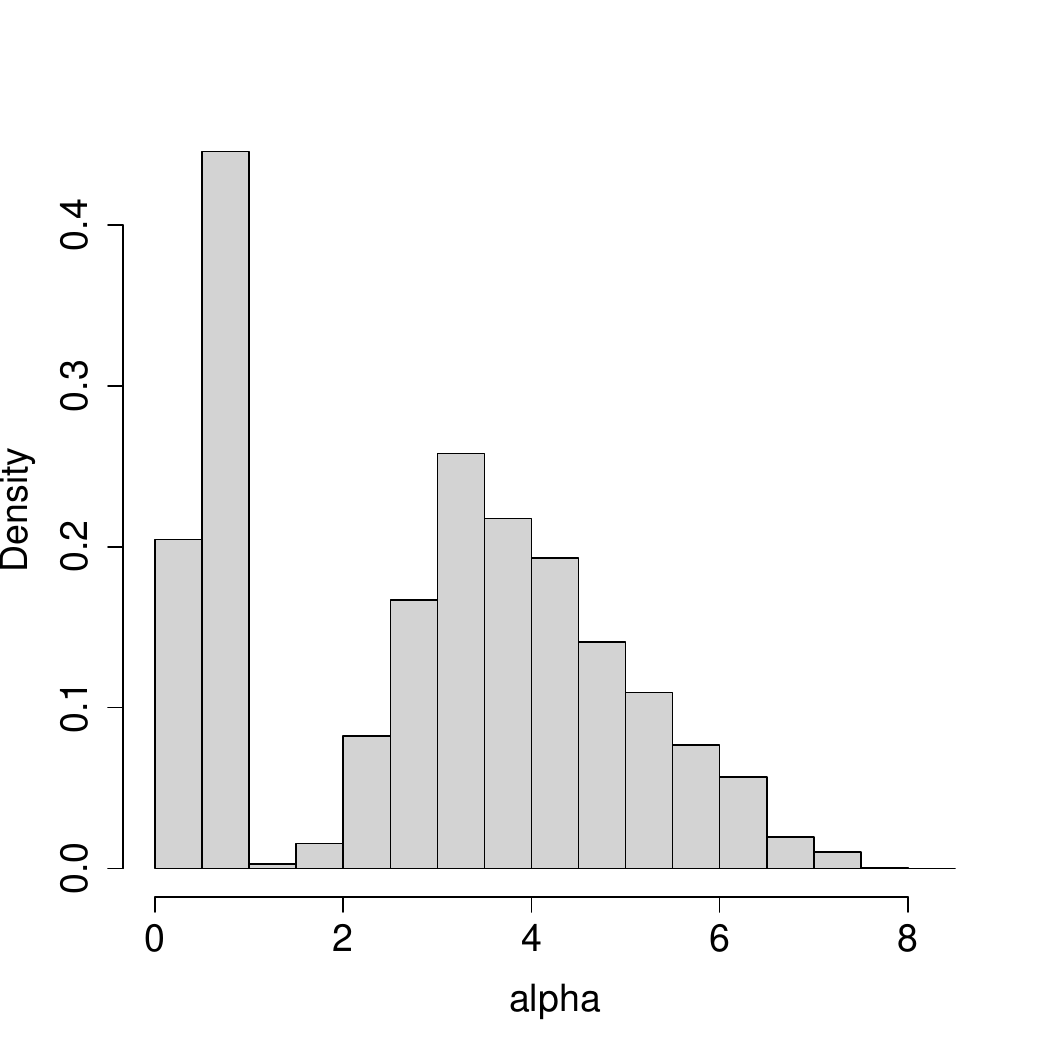}}
\caption{{\small Simulated data. Histogram of aggregated posterior samples from $\mu_i$ (left) and $\alpha_i$ (right), $i=1,\ldots,n$.}}
\label{fig:pars}
\end{figure}

The two most important parameters in the SGG model are $\mu$, which determines the location of a component, and $\alpha$, which determines the heaviness of the tail. Figure \ref{fig:pars} presents the histogram of the posterior distribution of these parameters, $\mu_i$ (left panel) and $\alpha_i$ (right panel) for $i=1,\ldots,n$. The posterior values of $\mu_i$ concentrate around the true values, $0$ and $5$. In addition, the posterior values of $\alpha_i$ concentrate around the true values, $0.5$ and $3$. This confirms that our posterior inference procedure is capable of reproducing the true model. 

\begin{figure}
\centerline{\includegraphics[scale=0.9]{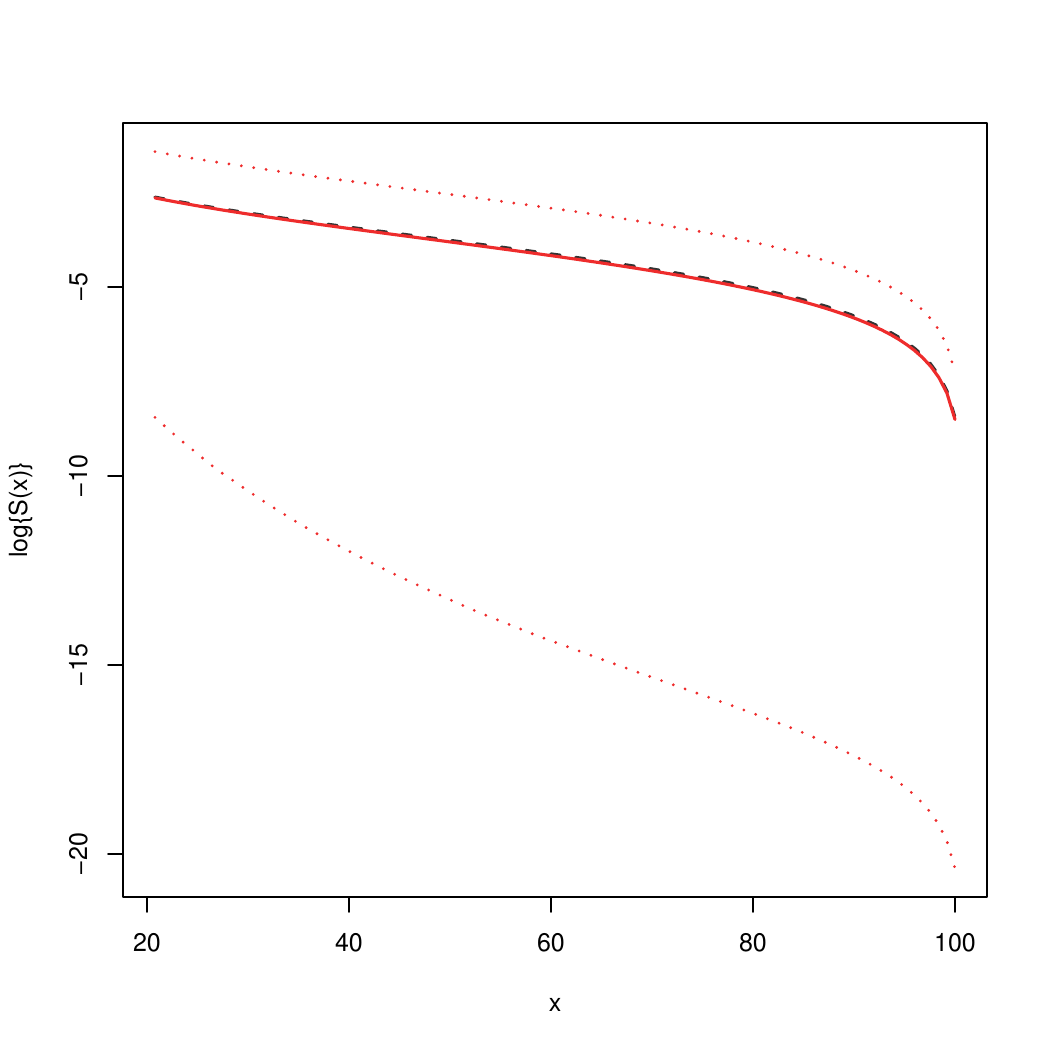}}
\caption{{\small Simulated data. Log survival function for $x>20$. True function (dashed line), posterior estimate (solid line), 95\% posterior CI (dotted lines). True and estimate are overlapped.}}
\label{fig:lsurvs}
\end{figure}

Figure \ref{fig:dens} includes the density estimates (posterior predictive density) as point (posterior mean) and $95\%$ credible intervals (CI). Our density estimate (solid line) closely follows the true density (dashed line) as well as the shape of the histogram of the data. When modelling heavy tail data, interest is on the tails. In Figure \ref{fig:lsurvs} we include a graph of the survival function $S(x)$ for large $x$ on a logarithmic scale. The estimated tail and the true tail are almost indistinguishable.

\subsection{Accidents insurance claims}

Federal Roads and Bridges of Mexico are administered by an agency called CAPUFE. This agency offers, among other benefits, an accident insurance policy to every car that uses a particular motorway in exchange for a fare. The insurance is covered by an insurance company that usually changes every two years. The available data consists of claims, in mexican pesos, filed with the insurance company during the month of June 2012. 

The size of the data is $n=1383$ with the following descriptive statistics: the minimum claim is $\$80$, the first quartile is $\$2,587$, the median is $\$6,000$, the third quartile is $\$15,790$ and the maximum is $\$790,107$. The mean is $\$17,158$, which is higher than the third quartile. These statistics suggest that the data might be heavy tailed. We therefore fit our mixture model  and to avoid numerical problems, the data were divided by $1,000$. 

The prior specifications for our model are the same as in the simulation study for all parameters and the MCMC was also defined as in the simulation study. Running times are reported in the last column of Table \ref{tab:fita}. Since the sample size is almost triple that of the simulation study and the number of groups required is also a lot larger, the running times increased considerably. These range from $29$ to $34$ minutes.

The three fit measures LPML, AIC and BIC are reported in Table \ref{tab:fita}. The smallest number of groups, $\widehat{m}=10$, is obtained when fixing $\kappa=0$ (NS case). Running time is the fastest, but the fit measures say that this is the worst model for the data. On the other hand, when we fix $\nu=0$ (DP case) and take the hyper prior $\kappa\sim\ga(1,2)$, the mode number of groups is the largest, $\widehat{m}=14$, but the three fit measures prefer this model. The running time is closely related to he number of groups $m$ in the mixture and the sample size, so fitting this winning model to these data takes around 34 minutes. The posterior mean of $\kappa$ is $1.75$ with a 95\% CI $[0.9,2.8]$. 

\begin{figure}
\centerline{\includegraphics[scale=0.45]{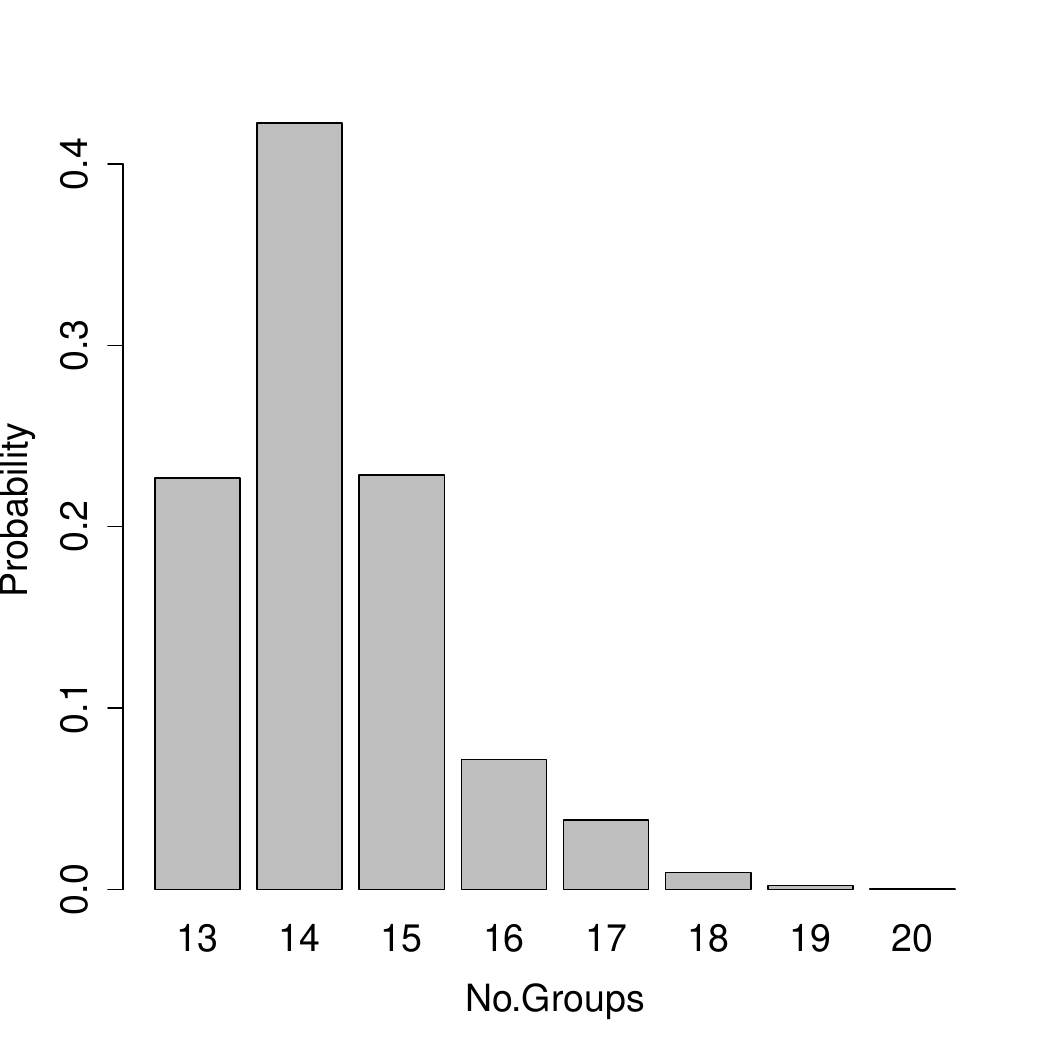}
\includegraphics[scale=0.45]{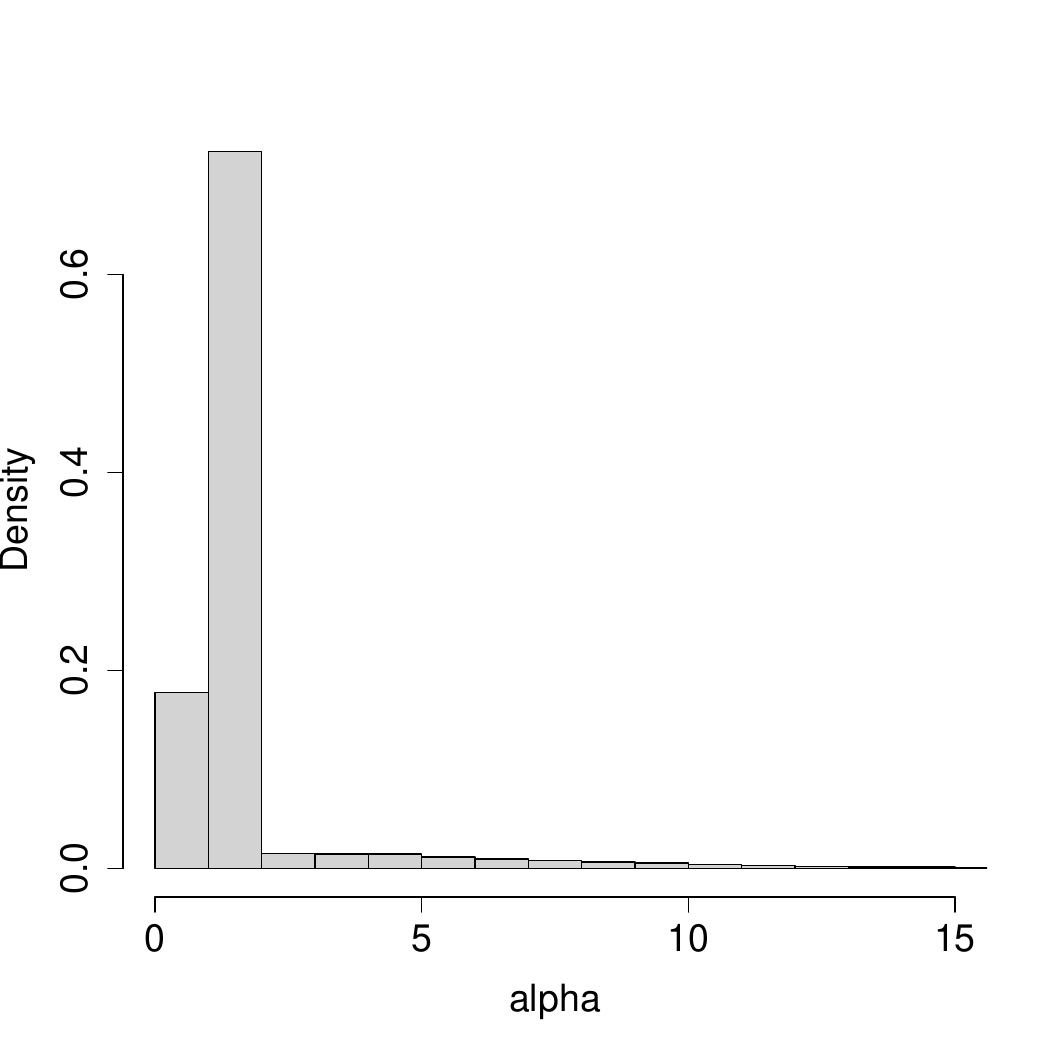}}
\caption{{\small Claims data. Posterior distribution of the number of groups $m$ (left) and aggregated $\alpha_i$, $i=1,\ldots,n$ (right).}}
\label{fig:n0a}
\end{figure}

The complete posterior distribution for the number of groups $m$ is presented in the left panel of Figure \ref{fig:n0a}. The mode is $14$ and the range goes from $13$ to $20$ groups. One of the most important parameters of our model is $\alpha$ because it determines the heaviness of the tails in the data. The posterior values of $\alpha_i$ were all aggregated for $i=1,\ldots,n$ and are shown in the right panel of Figure \ref{fig:n0a}. Most of the values are less than two, specifically $\P(\alpha<1)=0.18$, $\P(1\leq\alpha<2)=0.72$ and $\P(\alpha>2)=0.10$. We can say that $18\%$ of the data comes from a heavier tail component, $72\%$ of the data comes from a less heavy component with finite mean and infinite variance, and $10\%$ comes from a component with finite mean and variance. 

\begin{figure}
\centerline{\includegraphics[scale=0.9]{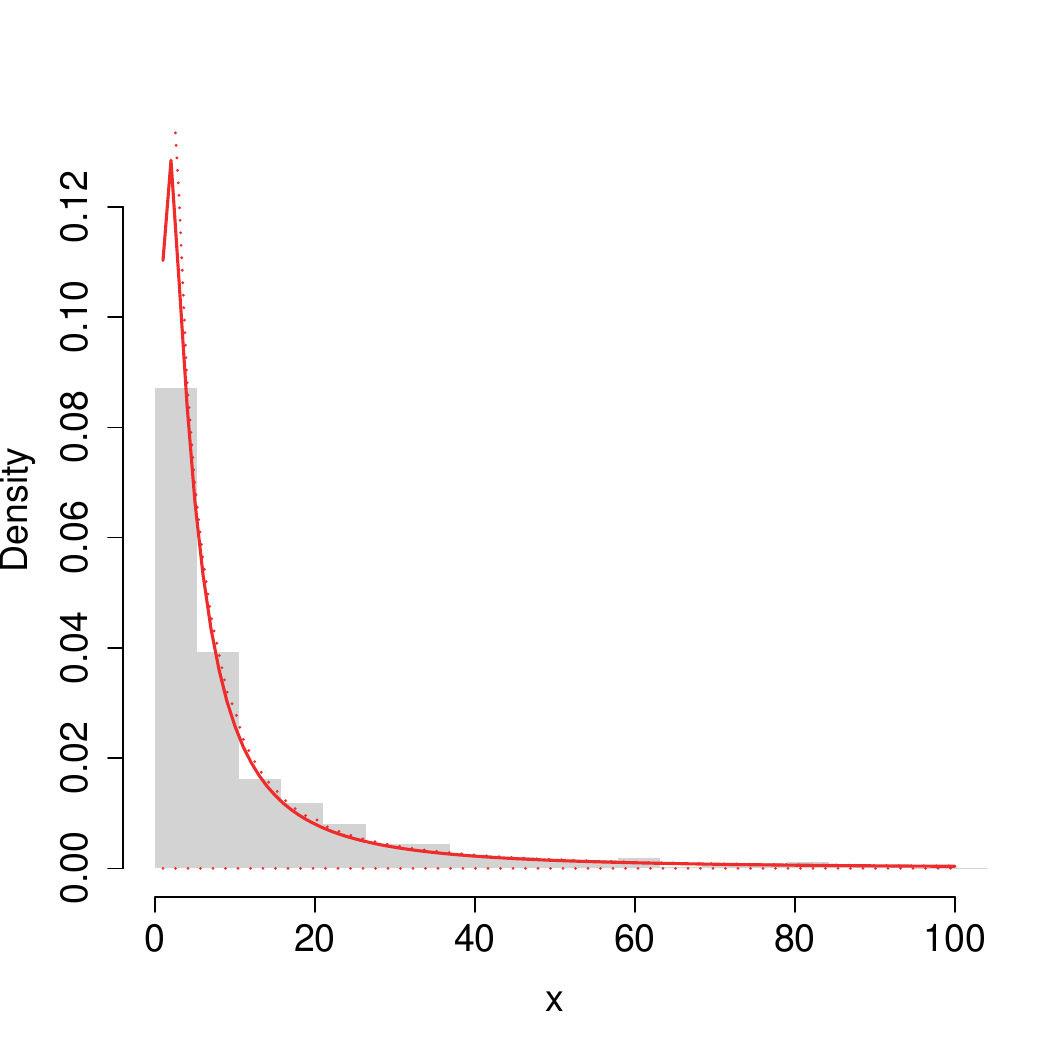}}
\caption{{\small Claims data. Histogram of data (background) and posterior density estimate (solid line), 95\% posterior CI (dotted lines).}}
\label{fig:dena}
\end{figure}

We finally obtain the density estimate via the posterior predictive distribution. This is shown in Figure \ref{fig:dena} together with $95\%$ CI. Our estimate closely follows the shape of the histogram, which was truncated to $[0,100]$ thousands, for visualisation purposes. The estimated probability of receiving a claim larger than $100$ thousand pesos is $0.045$.

To put our model in perspective, we also fitted a single SGG model \eqref{eq:sgg} with prior distributions for the model parameters given by $g_0(\theta)$ in \eqref{eq:g0}. To specify the prior, we took $a_{\mu}=0$, $b_{\mu}=100$, $a_{\gamma}=b_{\gamma}=1/2$, $a_{\alpha}=b_{\alpha}=1/2$ and $a_{\beta}=b_{\beta}=1/2$. The MCMC was specified by the same number of iterations, burn in and thinning, as that described for the mixture model above. As expected, all fit measures indicate that a single SGG model is the worst option, although the running time is one tenth of the mixture models. The estimated tail parameter is $\widehat{\alpha}=1.11$, which is in agreement with the tail parameters obtained with our mixture model.

\subsection{Population size in England}

As a second real data analysis we consider the population data set included in the R-package \texttt{poweRlaw}. This consists of $n=535$ population sizes of cities and towns in England in 2001 \citep{arcuate&al:14}. The descriptive statistics for these data are: the minimum size is $10.9$, the first quartile is $5,372$, the median is $11,987$, the third quartile is $31,692$ and the maximum is $7,659,513$. The mean is $61,753.7$, which is higher than the third quartile. Again, these statistics suggest the data might be heavy tailed. Before fitting our model, we divided the data by $1,000$ to avoid numerical problems. 

The prior specifications for model parameters $\mu$, $\gamma$, $\alpha$, $\beta$, $\nu$ and $\kappa$, and the MCMC specifications were also the same as for the simulation study. Although we have kept the same number of iterations as in the previous analysis, given that the sample size is now smaller and the number of groups in the mixture is also low, the running times, reported in the last column of Table \ref{tab:fitp}, are all below $4$ minutes. 

Again, we compare model fit by computing the LPML, and expected AIC and BIC measures. These are reported in Table \ref{tab:fitp}. In this case, the three measures take very similar values for the three PY prior specifications, with a slight disagreement to order the second and third options. However, the three measures agree on selecting the DP choice as the best model.  

The posterior distribution for the number of groups $m$ is presented in the right panel in Figure \ref{fig:n0}. The posterior mode is $2$ with $\P(m=2)=0.62$ and a range that goes from $2$ to $5$ groups. The posterior mean of the precision parameter $\kappa$ in the DP prior is $0.32$ with a $95\%$ CI $[0.04,0.88]$. 

\begin{figure}
\centerline{\includegraphics[scale=0.45]{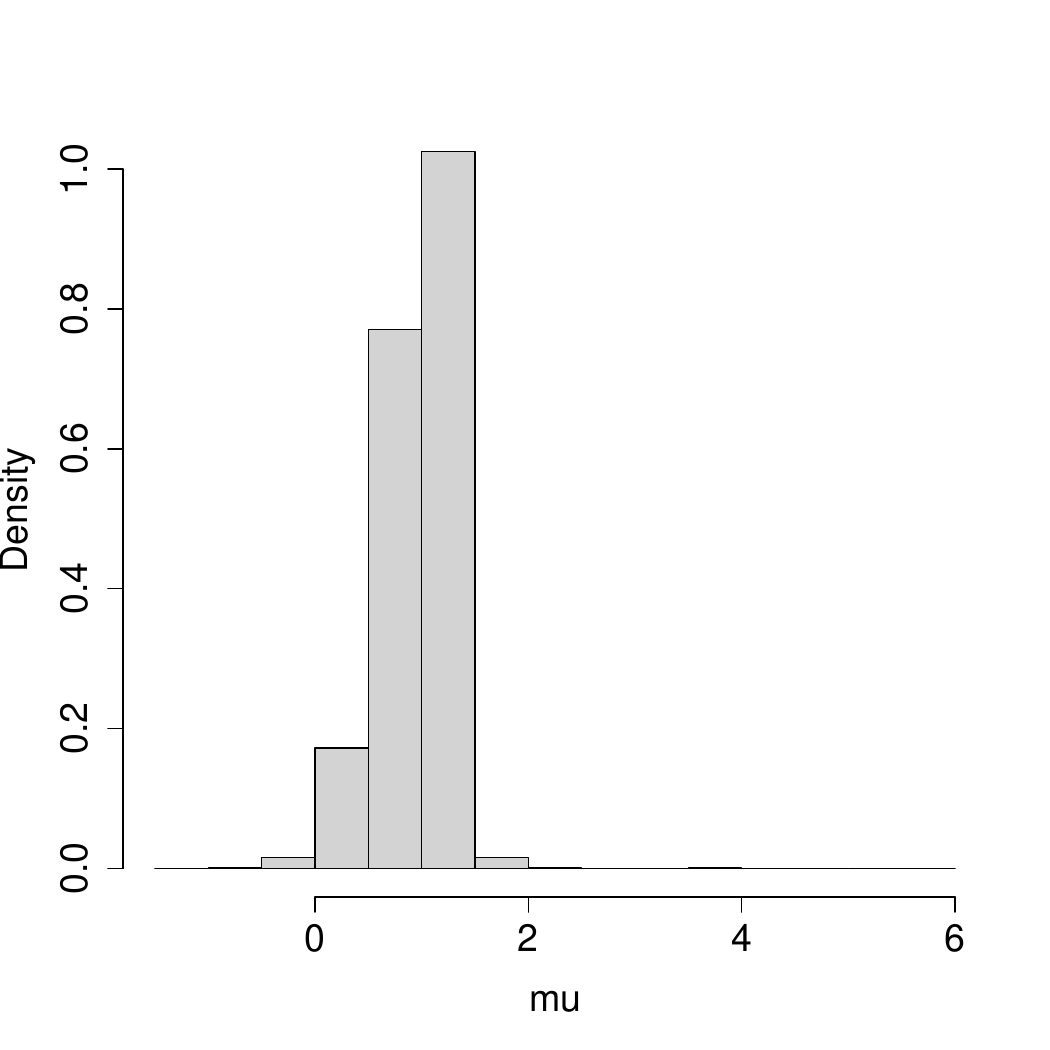}
\includegraphics[scale=0.45]{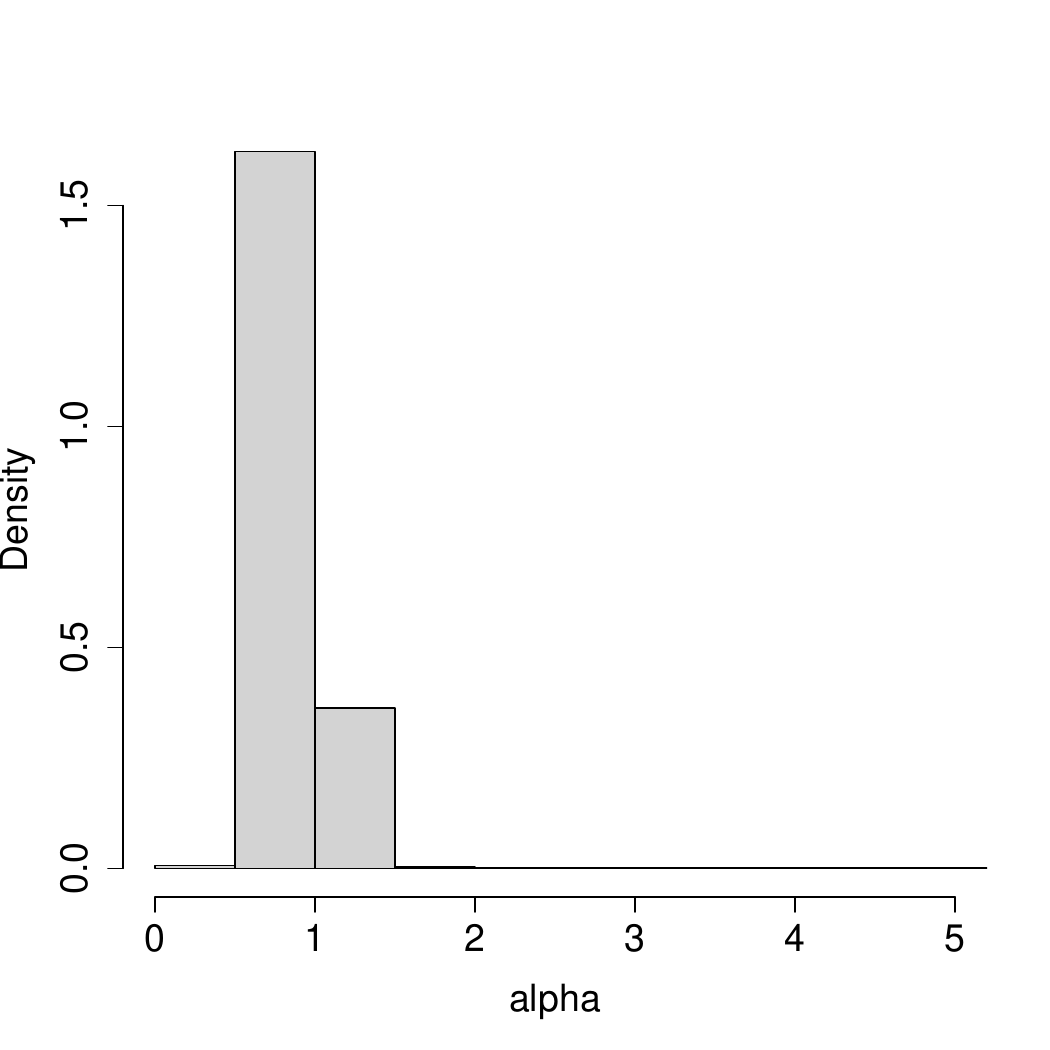}}
\caption{{\small Population data. Histogram of aggregated posterior samples from $\mu_i$ (left) and $\alpha_i$ (right), $i=1,\ldots,n$.}}
\label{fig:parp}
\end{figure}

The posterior distribution of parameters $\mu_i$ and $\alpha_i$, aggregated for all $i=1,\ldots,n$, are included in Figure \ref{fig:parp} in the left and right panel, respectively. Considering the location parameters, we obtain that $\P(\mu>0)=0.99$, which says that practically all locations are strictly positive. Moreover, since $\P(1/2<\mu<3/2)=0.9$, a great majority of them take values in the interval $(0.5,1.5)$. Now, considering the tail parameters, we get that $\P(\alpha<1)=0.81$, $\P(1\leq\alpha<2)=0.18$ and $\P(\alpha>2)=0.01$. We can say that $81\%$ of the data come from a heavier tailed component, and the rest $18\%$ of the data come from a less heavy component with finite mean but infinite variance. 

\begin{figure}
\centerline{\includegraphics[scale=0.9]{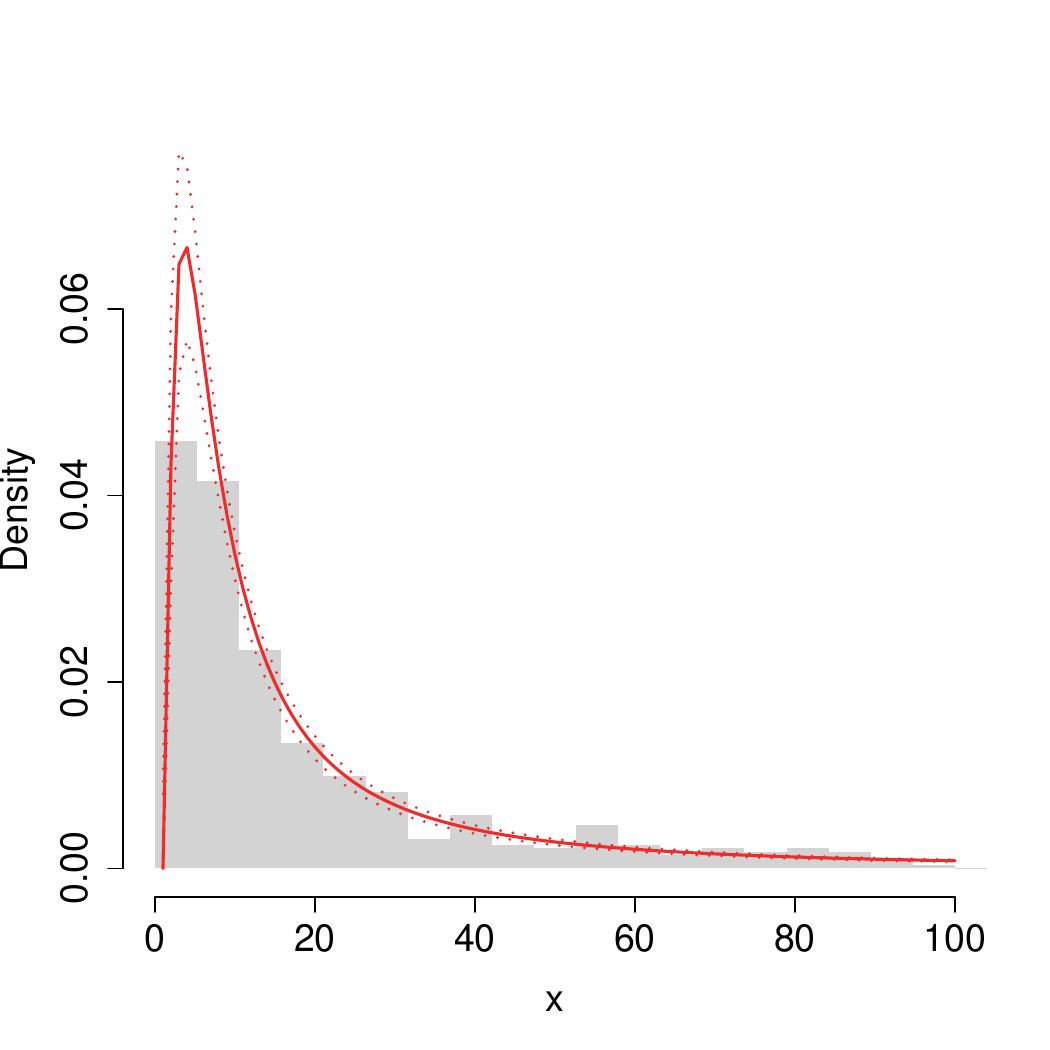}}
\caption{{\small Population data. Histogram of data (background) and posterior density estimate (solid line), 95\% posterior CI (dotted lines).}}
\label{fig:denp}
\end{figure}

We also produced the density estimate via the posterior predictive distribution. This is included in Figure \ref{fig:denp} as a solid line, together with $95\%$ CI as dotted lines. Our density estimate shows a large peak between $0$ and $10$ thousands, followed by a large right tail. Again, the graph was truncated to $[0,100]$ thousands, for visualisation purposes. The estimated probability of having a town/city with population larger than $100$ thousand habitants is $0.095$. 

Finally, to place our model in context, we also fitted a single SGG model with the same prior and MCMC specifications as in the previous data analysis. As expected, all fit measures indicate that a single SGG model is the worst option for these data. The estimated tail parameter is $\widehat{\alpha}=1.07$, which is also in agreement with the tail parameters obtained with our mixture model.

\section{Concluding remarks}
\label{sec:concl}

We have introduced a kernel based on the shifted gamma-gamma distribution. This is a very flexible model with four well identified parameters associated to location, shape, tail, and scale. In spite of the versatility of the kernel, we propose a Poisson-Dirichlet process as mixing measure over the four dimensional parameter vector to define a bayesian nonparametric mixture model for the analysis of heavy tail data.

Due to the form of the likelihood, there are no conjugate prior distributions for the model parameters. We managed to slightly simplify the likelihood by expressing it as a mixture of a conditional shifted gamma and a marginal gamma for the rate parameter. When combining the likelihood with the Poisson-Dirichlet process prior, we obtained a non conjugate marginal sampler which required a Metropolis-Hastings step. To optimise the proposal distribution, we introduced an efficient adapting mechanism that achieves a desire (optimal) acceptance rate. 

The code was implemented in Fortran in an Intel Xeon computer at 3.00 GHz and with 24GB of RAM. However, all calls to the executable file were made through the R Statistical Computing language. All codes as well as the datasets are available as Supplementary Material. 

Future work includes the inclusion of covariates in the analysis in a regression fashion and the study of mixtures for multivariate heavy tail data.

\section*{Funding}

This work was supported by \textit{Asociaci\'on Mexicana de Cultura, A.C.}

\section*{Conflicts of interest}

The author declares no conflict of interest. 

\section*{Data availability statement}

Data and codes that replicate the results of this study are available as Supplementary Material.

\bibliographystyle{natbib}

\newpage

\begin{table}
$$\begin{array}{ccccccc} 
\hline \hline
\nu & \kappa & \mbox{LPML} & \mbox{AIC} & \mbox{BIC} & \widehat{m} & \mbox{R.Time} \\ \hline
0 & \ga(1,2) & -1462 & 2741 & 2781 & 2 & 3.84 \\ 
\be(1/2,1/2) & 0 & -1727 & 2909 & 2950 & 2 & 3.88 \\ 
\be(1/2,1/2) & \ga(1,2) & -1524 & 2889 & 2932 & 2 & 3.69 \\ 
\hline \hline
\end{array}$$
\caption{Posterior summaries for simulated data with different values/priors for $(\nu,\kappa)$. Fit measures, number of groups (posterior mode) and running time (in minutes).}
\label{tab:fits}
\end{table}

\begin{table}
$$\begin{array}{ccccccc} 
\hline \hline
\nu & \kappa & \mbox{LPML} & \mbox{AIC} & \mbox{BIC} & \widehat{m} & \mbox{R.Time} \\ \hline
0 & \ga(1,2) & -2578 & 4524 & 4824 & 14 & 34.11 \\ 
\be(1/2,1/2) & 0 & -4234 & 7628 & 7852 & 10 & 29.11 \\ 
\be(1/2,1/2) & \ga(1,2) & -3002 & 5278 & 5533 & 12 & 31.77 \\ \hline
\multicolumn{2}{c}{\mbox{Single SGG}} & -4930 & 9865 & 9886 & - & 3.45 \\
\hline \hline
\end{array}$$
\caption{Posterior summaries for claims data with different values/prior for $(\nu.\kappa)$. Fit measures, number of groups (posterior mode) and running time (in minutes). Last row corresponds to the fit of a single SGG model.}
\label{tab:fita}
\end{table}

\begin{table}
$$\begin{array}{ccccccc} 
\hline \hline
\nu & \kappa & \mbox{LPML} & \mbox{AIC} & \mbox{BIC} & \widehat{m} & \mbox{R.Time} \\ \hline
0 & \ga(1,2) & -2306 & 4615 & 4657 & 2 & 3.75 \\ 
\be(1/2,1/2) & 0 & -2319 & 4617 & 4658 & 2 & 3.81 \\ 
\be(1/2,1/2) & \ga(1,2) & -2307 & 4616 & 4659 & 2 & 3.84 \\ \hline
\multicolumn{2}{c}{\mbox{Single SGG}} & -2342 & 4689 & 4706 & - & 1.35 \\
\hline \hline
\end{array}$$
\caption{Posterior summaries for population data with different value/prior for $(\nu,\kappa)$. Fit measures, number of groups (posterior mode) and running time (in minutes). Last row corresponds to the fit of a single SGG model.}
\label{tab:fitp}
\end{table}

\end{document}